\documentclass[11pt]{article}
\usepackage{a4,epsfig}
\usepackage{ifthen,array}
\newcommand{\ams}{\usepackage{amsfonts,amssymb,amsmath}}%,amsintx}}

\ams
\allowdisplaybreaks[3]
\newlength{\textwidthorig}
\newlength{\oddsidemarginorig}
\newlength{\textheightorig}
\newlength{\topmarginorig}
\def\seitenlaengenabsolut#1 #2 #3 #4 {\setlength{\textwidth}{#1}
                                      \setlength{\oddsidemargin}{#2}
                                      \setlength{\textheight}{#3}
                                      \setlength{\topmargin}{#4}}
\def\seitenlaengenrelzustandard#1 #2 #3 #4 {\setlength{\textwidth}{\textwidthorig+#1}
                                            \setlength{\oddsidemargin}{\oddsidemarginorig+#2}
                                            \setlength{\textheight}{\textheightorig+#3}
                                            \setlength{\topmargin}{\topmarginorig+#4}}
\def\seitenlaengenrelzuvorher#1 #2 #3 #4 {\addtolength{\textwidth}{#1}
                                          \addtolength{\oddsidemargin}{#2}
                                          \addtolength{\textheight}{#3}
                                          \addtolength{\topmargin}{#4}}
\newcommand{\standardseite}{\seitenlaengenrelzuvorher2.2cm -0.8cm 1.8cm -1.5cm }   %
\standardseite
\newcommand{\leerezeile}{\vspace{2ex}}
\newlength{\laengespatium}

\newcommand{\nach}{\longrightarrow}      %Abbildungspfeil

\newcommand{\auf}{\longmapsto}           %Abbildungspfeil fuer Elemente
\newcommand{\txtauf}[1]{\auf}            %Abbildungspfeil mit Opt. "Text dr"uber"
\newcommand{\impliz}{\Longrightarrow}    %Implikationspfeil
\newcommand{\invimpliz}{\Longleftarrow}  %Implikationspfeil (umgekehrte Richtung
\newcommand{\konvrunter}{\downarrow}     %Konvergenz von oben
\newcommand{\iso}{\cong}                 %Symbol fuer isomorph
\newcommand{\ident}{\equiv}              %Symbol fuer identisch (3 Striche)
\newcommand{\teilmenge}{\subseteq}       %Symbol fuer Teilmenge
\newcommand{\obermenge}{\supseteq}       %Symbol fuer Teilmenge andersherum
\newcommand{\echteteilmenge}{\subset}    %Symbol fuer echte Teilmenge
\newcommand{\aeqrel}{\sim}               %Symbol Aequivalenzrelation
\newcommand{\leeremenge}{\varnothing}    %Symbol f"ur leere Menge
\newcommand{\kreuz}{\times}              %Symbol f"ur Kreuz

\newcommand{\einschr}[1]{{}\arrowvert_{#1}}      %Einschr"ankung auf ...
\newcommand{\betraganpass}[1]%
           {\left| #1 \right|}           %Betragsstriche (variable Gr"o"se) 
\newcommand{\bigbetrag}[1]%
           {\bigl|{#1}\bigr|}            %Betragsstriche (gr"o"ser)
\newcommand{\betrag}[1]%
           {|{#1}|}                      %Betragsstriche  
\newcommand{\betragnichtanpass}[1]%
           {\mid #1 \mid}                %Betragsstriche 
\newcommand{\norm}[1]%
           {{}{\parallel}#1{\parallel}{}}      %Normstriche  
\newcommand{\erww}[1]%
           {\langle #1 \rangle}          %Erwartungwert-Klammern
\newcommand{\skalprod}[2]%
           {\langle #1,#2 \rangle}       %eckiges Skalarprodukt
\newcommand{\quer}{\overline}            %Strich drueber
\newcommand{\im}{\text{im\;}}                          %im als Image (mit platz)
\newcommand{\Ad}{{\text{Ad}}}                          %Ad als adjungiert
\newcommand{\elanz}{\#}                                %# f"ur Elementanzahl
\newcommand{\Hom}{\text{Hom}}                          %Hom als Homomorphismengruppe
\newcommand{\Maps}{\text{Maps}}                        %Maps f"ur Set of Maps
\newcommand{\field}[1]{\mathbb{#1}}                    %liefert #1 als mathbb-Zeichen
\newcommand{\C}{{\field{C}}}                           %C fuer komplexe Zahlen
\newcommand{\N}{{\field{N}}}                           %N fuer natuerliche Zahlen
\newcommand{\R}{{\field{R}}}                           %R fuer reelle Zahlen
\newcommand{\Gl}{\text{Gl}}                            %Lineare Gruppe
\newcommand{\gl}{\mathfrak{gl}}                        %Liealg. der Lin. Gruppe
\newcommand{\rnkl}[2]{\raisebox{-0.4ex}{$#1$}%
\raisebox{-0.12ex}{{\large$\setminus$}}\,#2}   %0.5/0.2 fr"uher  %Rechtsnebenklassenfaktorraum
\newcommand{\agb}{{\overline{{\cal A}/{\cal G}}}}      %A/G + Strich
\newcommand{\agbfact}[1][]{\text{$\agb/\!\aeqrel$}}    %A/G\~ + Strich
\newcommand{\ag}{{\cal A}/{\cal G}}                    %A/G ohne Strich
\newcommand{\Ab}{{\overline{{\cal A}}}}                %A + Strich
\newcommand{\A}{{\cal A}}                              %A ohne Strich
\newcommand{\Gb}{{\overline{{\cal G}}}}                %G + Strich

\newcommand{\G}{{\cal G}}                              %G ohne Strich
\newcommand{\AbGb}{{\Ab/\Gb}}                          %A+Strich / G+Strich

\newcommand{\abbAbGbagb}{{\boldsymbol \phi}}           %Abb. zwischen Ab/Gb und agb
\newcommand{\qa}{{\quer{A}}}                           %A als verallg. Zusammenhang 
\newcommand{\qg}{{\quer{g}}}                           %Verallgemeinerte Eichtrf.
\newcommand{\hg}{{\cal HG}}                            %HG
\newcommand{\holgr}{{\mathbf H}}                       %Holonomiegruppe
\newcommand{\bz}{{\mathbf B}}                          %Basiszentralisator
\newcommand{\Web}{{\text{Web}}}                        %Text "Web"
\newcommand{\GR}{\Gamma}                               %Graph Gamma
\newcommand{\Ver}{\mathbf{V}}                          %V fuer Vertexmenge
\newcommand{\gross}[1]{{\boldsymbol #1}}               %#1 als grosses
\newcommand{\gc}{\gross{\gamma}}                       %gro"ses gamma 
\newcommand{\pf}[1]{{\cal P}_{#1}}                     %Menge aller Pfade
\newcommand{\Pf}{{\cal P}}                             %Menge aller Pfade
\newcommand{\KG}[1]{\Pf_{#1}}                          %Wege_#1
\newcommand{\hyph}{\upsilon}                           %a hyph
\newcommand{\Haar}{{\text{Haar}}}                      %Index Haar
\newcommand{\LG}{{\mathbf{G}}}                         %Liegruppe G (fett)
\newcommand{\LN}{{\mathbf{N}}}                         %Liegruppe N (fett)
\newcommand{\Lieg}{{\mathfrak{g}}}                            %gotisches g (fuer Liealgebra)
\newcommand{\aeqrelzush}[1][]{\sim}                    %Symbol Aequivalenzrelation
\newcommand{\nklza}[1][]{\ifthenelse{\equal{#1}{}}     %Z(H_\qa) \ G
                                    {\rnkl{Z(\holgr_\qa)}{\LG}}        
                                   {\rnkl{Z(\holgr_{#1})}{\LG}}}       
\newcommand{\nkla}[1][]{\ifthenelse{\equal{#1}{}}      %B(\qa) \ \Gb
                                    {\rnkl{\bz(\qa)}{\Gb}}        
                                    {\rnkl{\bz(#1)}{\Gb}}}       
\newcommand{\etr}{g}                                   %klassische Eichtrf.
\newcommand{\he}{{\text{he}}}                          %he f"ur halbeinfach
\newcommand{\ab}{{\text{ab}}}                          %ab f"ur abelsch

\newcommand{\algnorm}[1]{\norm{#1}_\bullet}            %Algebranorm (fetter Punkt)
\newcommand{\FIL}[1]{\betrag{G_{#1}}}                  %Fl"acheninhalt zu Loop
\newcommand{\YM}{{\text{YM}}}                          %YM f"ur Yang-Mills

\newcommand{\ymwirk}[1][]{\ifthenelse{\equal{#1}{}}{S_{\YM}}{S_{\YM,#1}}}
\newcommand{\karte}{\chi}
\renewcommand{\karte}{{\kappa}}             %Karte

\newcommand{\bmat}{\begin{pmatrix}}
\newcommand{\emat}{\end{pmatrix}}

\newcommand{\const}{\text{const}}                      %konstant
\newcommand{\ListNullAbstaende}{\setlength{\topsep}{0pt}%
                                \setlength{\parskip}{0pt}%
                                \setlength{\partopsep}{0pt}%
                                \setlength{\itemsep}{0pt}%
                                \setlength{\parsep}{0pt}}
\newcommand{\ListNurAnstrichAbstand}{\setlength{\topsep}{0pt}%
                                     \setlength{\parskip}{0pt}%
                                     \setlength{\partopsep}{0pt}%
                                     \setlength{\parsep}{0pt}}
\newenvironment{StandardListe}[2]%
               {\begin{list}%
                      {#1}%
                      {\settowidth{\leftmargin}{M#1}%
                       \settowidth{\labelwidth}{#1}%
                       \settowidth{\labelsep}{M}%
                       #2%
                      }%
                }%
               {\end{list}}%
\newenvironment{EinfachListe}[1]%
               {\begin{StandardListe}{#1}{\ListNullAbstaende}}%
               {\end{StandardListe}}%
               {\begin{StandardListe}{#1}{\ListNurAnstrichAbstand}}%
               {\end{StandardListe}}%
\newcommand{\labelsatz}[1]{#1}
\newcounter{listennr}                      %
\newlength{\hilfslaenge}
\newlength{\stdlabellaenge}
\newlength{\maximum}
\newcommand{\stdlabel}{}
\newcommand{\Maximum}{}
\newcommand{\iitem}[1][]{\ifthenelse{\equal{#1}{}}%
                           {\item \setlength{\hilfslaenge}{\stdlabellaenge}}%
                           {\item[\labelsatz{#1}\hfill]%
                            \settowidth{\hilfslaenge}{\labelsatz{#1}}}%
                         \ifthenelse{\lengthtest{\maximum < \hilfslaenge}}%
                           {\setlength{\maximum}{\hilfslaenge}%
                            \ifthenelse{\equal{#1}{}}%
                               {\renewcommand{\Maximum}{\stdlabel}}%
                               {\renewcommand{\Maximum}{#1}}}%
                           {}%
                      }      
\makeatletter
\newenvironment{AutoLabelLaengenListe}[2][]%
               {\begin{list}%
                      {\labelsatz{#1}\hfill}%
                      {\stepcounter{listennr}%
                       \settowidth{\leftmargin}{M\labelsatz{\ref{listnr\arabic{listennr}}}}%
                       \settowidth{\labelwidth}{\labelsatz{\ref{listnr\arabic{listennr}}}}%
                       \settowidth{\labelsep}{M}%
                       \settowidth{\stdlabellaenge}{\labelsatz{#1}}%
                       \renewcommand{\stdlabel}{#1}%
                       #2%
                       \renewcommand{\Maximum}{}%
                      }%
                }%
               {\renewcommand{\@currentlabel}{\Maximum}%
                \label{listnr\arabic{listennr}}%
                \end{list}%
                }%
\makeatother
\newenvironment{StandardEinrueckung}[2]%
               {\begin{list}%
                      {#1}%
                      {\settowidth{\leftmargin}{M#1}%
                       \settowidth{\labelwidth}{#1}%
                       \settowidth{\labelsep}{M}%
                       #2%
                      }%
                \item}%
               {\end{list}}%
\newenvironment{Einrueckungpur}[1]%
               {\begin{StandardEinrueckung}{#1}{\ListNullAbstaende}}%
               {\end{StandardEinrueckung}}%
\newenvironment{Einrueckung}[1]%
               {\begin{StandardEinrueckung}{#1}{\setlength{\parsep}{0pt}}}%
               {\end{StandardEinrueckung}}%
\newcommand{\EineZeileGleichung}[2][0.0ex]
           {
            
            \vspace{#1} 
            \noindent
            \hspace*{\fill}
            $\displaystyle{#2}$
            \hspace*{\fill}

            \vspace{#1} 
            
           }

\makeatletter
\newcommand{\EineNumZeileGleichung}[2][0.5ex]
           {
            
            \vspace{#1} 
            \noindent
            \stepcounter{equation}
            \renewcommand{\@currentlabel}{\arabic{equation}}%
            \phantom{(\arabic{equation})}\hspace*{\fill}
            $\displaystyle{#2}$
            \hspace*{\fill}
            (\arabic{equation})

            \vspace{#1} 
            
           }
\makeatother
\makeatletter
\newcommand{\EineErwNumZeileGleichung}[2][0.5ex]
           {
            
            \vspace{#1} 
            \noindent
            \stepcounter{equation}
            \renewcommand{\@currentlabel}{\arabic{equation}}%
            \phantom{(\arabic{equation})}\hspace*{\fill}
            #2 %
            \hspace*{\fill}
            (\arabic{equation})

            \vspace{#1} 
            
           }
\makeatother

\newlength{\abstaug}              %
\newenvironment{AllgUnnumGleichung}[2][1.0ex]%           %
               {
  
                \setlength{\abstaug}{#1}
                \vspace{\abstaug}
                \hspace*{\fill}
                $\begin{array}[t]{#2}
                }%
               {\end{array}$
                \hspace*{\fill}
  
                \vspace{\abstaug}

                }%
\newenvironment{AllgNumGleichung}[2][0.0ex]%           %
               {
  
                \setlength{\abstaug}{#1}
                \vspace{\abstaug}
                $\begin{tabular*}{\textwidth}[t]{#2}
                }%
               {\end{tabular*}$

                \vspace{\abstaug}

               }%
\newenvironment{StandardUnnumGleichungKlein}[1][0ex]%       %
               {\renewcommand{\s}{\\[#1] }%
                \begin{AllgUnnumGleichung}{rcl}}%
               {\end{AllgUnnumGleichung}}%
\newenvironment{StandardUnnumGleichung}[1][0ex]%       
               {\renewcommand{\s}{\\[#1] }%
                \begin{AllgUnnumGleichung}{>{\displaystyle}rc>{\displaystyle}l}}%
               {\end{AllgUnnumGleichung}}%
\newenvironment{XrelYZNumGleichung}[1][0ex]%       %
               {\renewcommand{\s}{\\[#1] }%
                \begin{AllgNumGleichung}{rcll}}%
               {\end{AllgNumGleichung}}%

\newcommand{\erllang}[2][0.5\textwidth]%
              {\hfill\hspace*{1.5em}%
               \begin{minipage}[t]{#1}{\small%
                          \begin{list}{(}{\ListNullAbstaende%
                                          \settowidth{\leftmargin}{(}%
                                          \settowidth{\labelwidth}{(}%
                                          \settowidth{\labelsep}{}%
                                         }%
                          \item#2)%
                          \end{list}}%
               \end{minipage}\\[-0.9ex]
              }%         
\newcommand{\DefBemUmgeb}[1]% 
           {\newenvironment{#1}[1][]%
                           {\begin{Einrueckung}{{\bf #1}}%
                            \ifx##1\empty\else{{\bf ##1}
                            
                                                        }\fi%
                            }%
                           {\end{Einrueckung}}}
\newcommand{\DefSBemUmgeb}[2]% %
           {\newenvironment{#1}[1][]%
                           {\begin{Einrueckung}{{\bf #2}}%
                            \ifx##1\empty\else{{\bf ##1}
                            
                                                        }\fi%
                            }%
                           {\end{Einrueckung}}}
\makeatletter
\newcommand{\DefBspUmgeb}[3]% %
           {\newcounter{#2}[#3]%
            \newenvironment{#1}[1][]%
                           {\stepcounter{#2}%
                            \renewcommand{\ZaehlerMarke}{\arabic{#2}}%  
                            \renewcommand{\Einzugsname}{{\bf #1 \ZaehlerMarke}}%
                            \begin{Einrueckung}{\Einzugsname}
                            \ifx##1\empty\else{{\bf ##1}\\}\fi%
                            \renewcommand{\@currentlabel}{\ZaehlerMarke}%
                            }%
                           {\end{Einrueckung}}}
\makeatother
\newcommand{\ZaehlerbisEbene}{section}
\newcommand{\Ebenea}{section}
\newcommand{\Ebeneb}{subsection}

\newcommand{\Abschnittnummer}{%
            \ifx\ZaehlerbisEbene\Ebenea{\arabic{section}}%
             \else{%
              \ifx\ZaehlerbisEbene\Ebeneb{\arabic{section}.\arabic{subsection}}%
               \else{\arabic{section}.\arabic{subsection}.\arabic{subsubsection}}%
              \fi}%     
            \fi}     
\newcommand{\Abschnittnummerpunkt}{\Abschnittnummer.}     %  {}%
\newcommand{\Einzugsname}{}
\newcommand{\ZaehlerMarke}{}
\makeatletter
\newcommand{\DefThmUmgeb}[3]% 
           {\newcounter{#1}[#3]%
            \newenvironment{#1}[1][]%
                           {\stepcounter{#2}%
                            \setcounter{#1}{\value{#2}}%
                            \renewcommand{\ZaehlerMarke}{\Abschnittnummerpunkt\arabic{#1}}%  
                            \renewcommand{\Einzugsname}{{\bf #1 \ZaehlerMarke}}%
                            \begin{Einrueckung}{\Einzugsname}
                            \ifx##1\empty\else{{\bf ##1}
                            
                                                        }\fi%
                            \renewcommand{\@currentlabel}{\ZaehlerMarke}%
                            }%
                           {\end{Einrueckung}}}
\makeatother
\makeatletter
\newcommand{\DefSThmUmgeb}[4]% 
           {\newcounter{#1}[#3]%
            \newenvironment{#1}[1][]%
                           {\stepcounter{#2}%
                            \setcounter{#1}{\value{#2}}%
                            \renewcommand{\ZaehlerMarke}{\Abschnittnummerpunkt\arabic{#1}}%
                            \renewcommand{\Einzugsname}{{\bf #4 \ZaehlerMarke}}
                            \begin{Einrueckung}{\Einzugsname}
                            \ifx##1\empty\else{{\bf ##1}

                                                        }\fi%
                            \renewcommand{\@currentlabel}{\ZaehlerMarke}%
                            }%
                           {\end{Einrueckung}}}
\makeatother
\makeatletter
\newcommand{\DefUnterNumThmUmgeb}[5]% 
           {\newcounter{#1}[#3]%
            \newcounter{#4}%
            \newenvironment{#1}[1][]%
                           {\ifx##1\empty\else{\stepcounter{#2}\setcounter{#4}{0}}\fi%
                            \stepcounter{#4}%
                            \setcounter{#1}{\value{#2}}%
                            \renewcommand{\ZaehlerMarke}{\Abschnittnummerpunkt\arabic{#1}\alph{#4}}%
                            \renewcommand{\Einzugsname}{{\bf #5 \ZaehlerMarke}}
                            \begin{Einrueckung}{\Einzugsname}
                            \renewcommand{\@currentlabel}{\ZaehlerMarke}%
                            }%
                           {\end{Einrueckung}}}
\makeatother
\newenvironment{Beweis}[1][]%
               {\begin{Einrueckung}{{\bf Beweis}}%
                \ifx#1\empty\else{{\bf #1}

                                            }\fi%
                }%
               {\end{Einrueckung}%
                }%
\newenvironment{Proof}[1][]%
               {\begin{Einrueckung}{{\bf Proof}}%
                \ifx#1\empty\else{{\bf #1}

                                            }\fi%
                }%
               {\end{Einrueckung}%
                }%
               {\begin{Einrueckung}{{\bf \glqq Beweis\grqq}}%
                \ifx#1\empty\else{{\bf #1}
                
                                            }\fi%
                }%
               {\end{Einrueckung}%
                }%
               {\begin{Einrueckung}{{\bf Begr"undung}}%
                \ifx#1\empty\else{{\bf #1}
                
                                            }\fi%
                }%
               {\end{Einrueckung}%
                }%
\newenvironment{Hinrichtung}%
               {\begin{Einrueckungpur}{$\impliz$}}%
               {\end{Einrueckungpur}}%
\newenvironment{Rueckrichtung}%
               {\begin{Einrueckungpur}{$\invimpliz$}}%
               {\end{Einrueckungpur}}%
               {\begin{Einrueckungpur}{\glqq$\teilmenge$\grqq}}%
               {\end{Einrueckungpur}}%
               {\begin{Einrueckungpur}{\glqq$\obermenge$\grqq}}%
               {\end{Einrueckungpur}}%
               {\begin{Einrueckungpur}{"$\teilmenge$"}}%
               {\end{Einrueckungpur}}%
               {\begin{Einrueckungpur}{"$\obermenge$"}}%
               {\end{Einrueckungpur}}%
\newcommand{\qed}{\nopagebreak\hspace*{2em}\hspace*{\fill}{\bf qed}}
\newcommand{\ARabic}{\arabic}
\newcommand{\Nummerntypa}{\arabic}   
\newcommand{\Nummerntypb}{\alph}
\newcommand{\Nummerntypc}{\roman}
\newcommand{\Nummerntypd}{\Alph}

\newcommand{\Nra}{\Nummerntypa{Nummera}}            %
\newcommand{\Nrb}{\Nummerntypb{Nummerb}}            %
\newcommand{\Nrc}{\Nummerntypc{Nummerc}}                
\newcommand{\Nrd}{\Nummerntypd{Nummerd}}                
\newcommand{\ZeichenzuNrTyp}[1]%
           {\ifx#1\ARabic {.}\else{)}%
                  \fi}                              %
\newcommand{\NrZeicha}{\ZeichenzuNrTyp{\Nummerntypa}}
\newcommand{\NrZeichb}{\ZeichenzuNrTyp{\Nummerntypb}}
\newcommand{\NrZeichc}{\ZeichenzuNrTyp{\Nummerntypc}}
\newcommand{\NrZeichd}{\ZeichenzuNrTyp{\Nummerntypd}}
\newcommand{\ListMarkea}%
           {\Nra\NrZeicha}
\newcommand{\ListMarkeb}%
           {\Nra\NrZeicha\Nrb\NrZeichb}
\newcommand{\ListMarkec}%
           {\Nra\NrZeicha\Nrb\NrZeichb\Nrc\NrZeichc}
\newcommand{\ListMarked}%
           {\Nra\NrZeicha\Nrb\NrZeichb\Nrc\NrZeichc\Nrd\NrZeichd}
\newcommand{\Anfangszeichen}{}
\newcommand{\Anfangspunkt}{}
\newcounter{Schachtelebene}
\newcounter{Hilfszaehler}
\newcommand{\Hilfsbefehl}{}
\newcommand{\Schachtelebene}{\alph{Schachtelebene}}
\makeatletter
\newenvironment{AllgNumerierteListe}[2][]%      %
               {\addtocounter{Schachtelebene}{1}%
		\setcounter{Hilfszaehler}{#2}%
                \renewcommand{\Anfangszeichen}%
                             {\renewcommand{\Hilfsbefehl}{\csname Nummerntyp\Schachtelebene \endcsname}%
                              \Hilfsbefehl{Hilfszaehler}}%
                \renewcommand{\Anfangspunkt}%
                             {\csname NrZeich\Schachtelebene \endcsname}%
                \begin{list}%
                      {\stepcounter{Nummer\Schachtelebene}%
                       \csname Nr\Schachtelebene \endcsname
                       \csname NrZeich\Schachtelebene \endcsname
                       }%
                      {\settowidth{\leftmargin}{M\Anfangszeichen\Anfangspunkt}%
                       \settowidth{\labelwidth}{\Anfangszeichen\Anfangspunkt}%
                       \settowidth{\labelsep}{M}%
                       \setlength{\topsep}{0pt}%
                       \setlength{\parskip}{0pt}%
                       \setlength{\partopsep}{0pt}%
                       \setlength{\itemsep}{0pt}%
                       \setlength{\parsep}{0pt}%
                      }%
                \renewcommand{\@currentlabel}{\csname ListMarke\Schachtelebene \endcsname}%
                }%      
               {\ifthenelse{\equal{}{}}{\setcounter{Nummer\Schachtelebene}{0}}{}
                \addtocounter{Schachtelebene}{-1}%
                \end{list}}
\makeatother
\newenvironment{NumerierteListe}[1]%      %
               {\begin{AllgNumerierteListe}{#1}}
               {\end{AllgNumerierteListe}}
\newenvironment{WeiterNumerierteListe}[1]%      %
               {\begin{AllgNumerierteListe}[Weiter]{#1}}
               {\end{AllgNumerierteListe}}

\newcommand{\UnnumAnfangszeichen}{}
\newcounter{UnnumSchachtelebene}
\newcommand{\UnnumSchachtelebene}{\alph{UnnumSchachtelebene}}
\makeatletter
\newenvironment{UnnumerierteListe}%          
               {\addtocounter{UnnumSchachtelebene}{1}%
                \renewcommand{\UnnumAnfangszeichen}%
                             {\csname UnnumZeich\UnnumSchachtelebene \endcsname}%
                \begin{list}%
                      {\UnnumAnfangszeichen}%
                      {\settowidth{\leftmargin}{M\UnnumAnfangszeichen}%
                       \settowidth{\labelwidth}{\UnnumAnfangszeichen}%
                       \settowidth{\labelsep}{M}%
                       \setlength{\topsep}{0pt}%
                       \setlength{\parskip}{0pt}%
                       \setlength{\partopsep}{0pt}%
                       \setlength{\itemsep}{0pt}%
                       \setlength{\parsep}{0pt}%
                      }%
                }%
               {\addtocounter{UnnumSchachtelebene}{-1}%
                \end{list}}
\makeatother
\newlength{\fktdefhilfslaenge}
\newcommand{\ohnefktdef}[4]%                 %
           {\hspace*{\fill}
            $\begin{array}[t]{ccc}%
            #1 & \nach & #2 \\
            #3 & \auf  & #4
            \end{array}$
            \hspace*{\fill}}
\newcommand{\fktdef}[5]%                 %
           {\hspace*{\fill}
            $\begin{array}[t]{cccc}%
            #1: & #2 & \nach & #3 \\    
                & #4 & \auf  & #5
            \end{array}$
            \settowidth{\fktdefhilfslaenge}{$#1$:}
            \hspace*{0.6 \fktdefhilfslaenge}  
            \hspace*{\fill}}
\newcommand{\fktdefpur}[5]%                 %
           {$\begin{array}[t]{cccc}%
            #1: & #2 & \nach & #3 \\    
                & #4 & \auf  & #5
            \end{array}$}
\newcommand{\fktdefabgesetztpur}[5]%          %
           {
            
            $\begin{array}[t]{cccc}%
            #1: & #2 & \nach & #3 \\    
                & #4 & \auf  & #5
            \end{array}$
            \settowidth{\fktdefhilfslaenge}{$#1$:}
            \hspace*{0.6 \fktdefhilfslaenge}
            
           }
\newcommand{\fktdefabgesetzt}[5]%                %
           {
           
            \hspace*{\fill}
            $\begin{array}[t]{cccc}%
            #1: & #2 & \nach & #3 \\    
                & #4 & \auf  & #5
            \end{array}$
            \settowidth{\fktdefhilfslaenge}{$#1$:}
            \hspace*{0.6 \fktdefhilfslaenge}  
            \hspace*{\fill}
            
            }
\newcommand{\ohnefktdefabgesetzt}[4]%                %
           {      

            \hspace*{\fill}
            $\begin{array}[t]{ccc}%
            #1 & \nach & #2 \\
            #3 & \auf  & #4
            \end{array}$
            \hspace*{\fill}

            }
\newcommand{\doppelohnefktdefabgesetzt}[6]%                %
           {

            \hspace*{\fill}
            $\begin{array}[t]{ccccc}%
            #1 & \nach & #2 & \nach & #3\\
            #4 & \auf  & #5 & \auf  & #6
            \end{array}$
            \hspace*{\fill}

            }
\newcommand{\anhang}%
           {\appendix
            \sectioninh{Anhang}
            \renewcommand{\Abschnittnummer}{%
                  \ifx\ZaehlerbisEbene\Ebenea{\Alph{section}}%
                  \else{%
                        \ifx\ZaehlerbisEbene\Ebeneb{\Alph{section}.\arabic{subsection}}%
                        \else{\Alph{section}.\arabic{subsection}.\arabic{subsubsection}}%
                        \fi}%     
                  \fi}%
            \renewcommand{\Abschnittnummerpunkt}{\Abschnittnummer.}     
            }            
\newcommand{\anhangengl}%
           {\appendix
            \sectioninh{Appendix}
            \renewcommand{\Abschnittnummer}{%
                  \ifx\ZaehlerbisEbene\Ebenea{\Alph{section}}%
                  \else{%
                        \ifx\ZaehlerbisEbene\Ebeneb{\Alph{section}.\arabic{subsection}}%
                        \else{\Alph{section}.\arabic{subsection}.\arabic{subsubsection}}%
                        \fi}%     
                  \fi}%
            \renewcommand{\Abschnittnummerpunkt}{\Abschnittnummer.}     
            }

\newcounter{wdhlstufe}
\newcommand{\sectioninh}[1]%
           {\section*{#1}%
            \addcontentsline{toc}{section}{#1}}
\newcommand{\bezeichnung}[3]%
           {\begin{Einrueckungpur}{\hbox to 6em{#1}\hbox to 2.4em{\hfill#2}}
            #3
            \end{Einrueckungpur}}

\newcommand{\doppelteinfach}{e}

\newcommand{\ifdoppelt}[1]{\ifthenelse{\equal{\doppelteinfach}{d}}{#1}{}}
\newcommand{\ifeinfach}[1]{\ifthenelse{\equal{\doppelteinfach}{e}}{#1}{}}

\newlength{\querfhilfsl}              %

\newlength{\hll}

\DefThmUmgeb{Theorem}{Theorem}{\ZaehlerbisEbene}
\DefThmUmgeb{Definition}{Definition}{\ZaehlerbisEbene}
\DefThmUmgeb{Satz}{Theorem}{\ZaehlerbisEbene}
\DefThmUmgeb{Proposition}{Theorem}{\ZaehlerbisEbene}
\DefThmUmgeb{Lemma}{Theorem}{\ZaehlerbisEbene}
\DefThmUmgeb{Folgerung}{Theorem}{\ZaehlerbisEbene}
\DefThmUmgeb{Corollary}{Theorem}{\ZaehlerbisEbene}
\DefThmUmgeb{Vorschrift}{Definition}{\ZaehlerbisEbene}
\DefThmUmgeb{Construction}{Definition}{\ZaehlerbisEbene}
\DefSThmUmgeb{FormSatz}{Theorem}{\ZaehlerbisEbene}{\glqq Satz\grqq} 
\DefThmUmgeb{Vermutung}{Theorem}{\ZaehlerbisEbene}
\DefThmUmgeb{Conjecture}{Theorem}{\ZaehlerbisEbene}
\DefThmUmgeb{Konvention}{Definition}{\ZaehlerbisEbene}
\DefThmUmgeb{Feststellung}{Theorem}{\ZaehlerbisEbene}
\DefUnterNumThmUmgeb{DefinitionZusatzNum}{Definition}{\ZaehlerbisEbene}{DefZN}{Definition}
\DefBspUmgeb{Beispiel}{Beispiel}{subsubsection}
\DefBspUmgeb{Example}{Example}{subsubsection}
\DefBspUmgeb{Frage}{Frage}{section}
\DefBspUmgeb{Question}{Question}{section}
\DefBspUmgeb{Aufgabe}{Aufgabe}{section}
\DefBspUmgeb{Ziel}{Ziel}{section}
\DefBemUmgeb{Bemerkung}
\DefBemUmgeb{Remark}
\DefSBemUmgeb{OffeneFrage}{Offene Frage}
\newcommand{\bdf}{\begin{Definition}}
\newcommand{\edf}{\end{Definition}}
\newcommand{\bvorsch}{\begin{Vorschrift}}
\newcommand{\evorsch}{\end{Vorschrift}}
\newcommand{\bconst}{\begin{Construction}}
\newcommand{\econst}{\end{Construction}}
\newcommand{\bthm}{\begin{Theorem}}
\newcommand{\ethm}{\end{Theorem}}
\newcommand{\bsatz}{\begin{Satz}}
\newcommand{\esatz}{\end{Satz}}
\newcommand{\bprop}{\begin{Proposition}}
\newcommand{\eprop}{\end{Proposition}}
\newcommand{\blem}{\begin{Lemma}}
\newcommand{\elem}{\end{Lemma}}
\newcommand{\bfolg}{\begin{Folgerung}}
\newcommand{\efolg}{\end{Folgerung}}
\newcommand{\bcorr}{\begin{Corollary}}
\newcommand{\ecorr}{\end{Corollary}}
\newcommand{\bfest}{\begin{Feststellung}}
\newcommand{\efest}{\end{Feststellung}}
\newcommand{\bbew}{\begin{Beweis}}
\newcommand{\ebew}{\end{Beweis}}
\newcommand{\bpf}{\begin{Proof}}
\newcommand{\epf}{\end{Proof}}
\newcommand{\bwnum}{\begin{WeiterNumerierteListe}}
\newcommand{\ewnum}{\end{WeiterNumerierteListe}}
\newcommand{\bdfzn}{\begin{DefinitionZusatzNum}}
\newcommand{\edfzn}{\end{DefinitionZusatzNum}}
\newcommand{\bbem}{\begin{Bemerkung}}
\newcommand{\ebem}{\end{Bemerkung}}
\newcommand{\brem}{\begin{Remark}}
\newcommand{\erem}{\end{Remark}}
\newcommand{\bnum}{\begin{NumerierteListe}}
\newcommand{\enum}{\end{NumerierteListe}}
\newcommand{\bunum}{\begin{UnnumerierteListe}}
\newcommand{\eunum}{\end{UnnumerierteListe}}
\newcommand{\bbsp}{\begin{Beispiel}}
\newcommand{\ebsp}{\end{Beispiel}}
\newcommand{\bex}{\begin{Example}}
\newcommand{\eex}{\end{Example}}
\newcommand{\bfrag}{\begin{Frage}}
\newcommand{\efrag}{\end{Frage}}
\newcommand{\bquest}{\begin{Question}}
\newcommand{\equest}{\end{Question}}
\newcommand{\baufg}{\begin{Aufgabe}}
\newcommand{\eaufg}{\end{Aufgabe}}
\newcommand{\bof}{\begin{OffeneFrage}}
\newcommand{\eof}{\end{OffeneFrage}}
\newcommand{\bverm}{\begin{Vermutung}}
\newcommand{\everm}{\end{Vermutung}}
\newcommand{\bconj}{\begin{Conjecture}}
\newcommand{\econj}{\end{Conjecture}}
\newcommand{\bkonv}{\begin{Konvention}}
\newcommand{\ekonv}{\end{Konvention}}
\newcommand{\bglklein}{\begin{StandardUnnumGleichungKlein}}
\newcommand{\eglklein}{\end{StandardUnnumGleichungKlein}}
\newcommand{\bgl}{\begin{StandardUnnumGleichung}}
\newcommand{\egl}{\end{StandardUnnumGleichung}}
\newcommand{\bglrtext}{\begin{XrelYZNumGleichung}}
\newcommand{\eglrtext}{\end{XrelYZNumGleichung}}
\newcommand{\zgl}{\EineZeileGleichung}

\newcommand{\zglklein}[1]{\zgl{\textstyle#1}}

\newcommand{\berlgl}{\begin{StandardUnnumGleichung}}
\newcommand{\eerlgl}{\end{StandardUnnumGleichung}}
\newcommand{\beinrueck}{\begin{Einrueckungpur}} 
\newcommand{\eeinrueck}{\end{Einrueckungpur}}
\newcommand{\beinflist}{\begin{EinfachListe}} 
\newcommand{\eeinflist}{\end{EinfachListe}}
\newcommand{\beq}{\begin{equation}}
\newcommand{\eeq}{\end{equation}}
\newcommand{\bhin}{\begin{Hinrichtung}}
\newcommand{\ehin}{\end{Hinrichtung}}
\newcommand{\brueck}{\begin{Rueckrichtung}}
\newcommand{\erueck}{\end{Rueckrichtung}}
\newcommand{\bvl}{\begin{AutoLabelLaengenListe}{\ListNullAbstaende}}
\newcommand{\evl}{\end{AutoLabelLaengenListe}}
\newcommand{\df}[1]{{\bf #1}}

\renewcommand{\he}{{\text{ss}}}                          
\newcommand{\partr}{\tau}
\newcommand{\spezabb}{\theta}
\newcommand{\pl}[1]{{\overline{\cal#1}}}
\newcommand{\eichtrf}{g}
\newcommand{\triv}{\Xi}
\newcommand{\ph}{^{\phantom{1}}}
\renewcommand{\he}{{\text{ss}}}                      
\newlength{\adressabstand}
\usepackage{doublespace}
\begin{document}
\begin{spacing}{1}
\title{{Regular Connections among Generalized Connections}}
\author{Christian Fleischhack\thanks{e-mail: 
            chfl@mis.mpg.de} \\   %
        \\
        {\normalsize\em Max-Planck-Institut f\"ur Mathematik in den
                        Naturwissenschaften}\\[\adressabstand]
        {\normalsize\em Inselstra\ss e 22--26}\\[\adressabstand]
        {\normalsize\em 04103 Leipzig, Germany}
        \\[-25\adressabstand]      %
        {\normalsize\em Center for Gravitational Physics and Geometry}\\[\adressabstand]
        {\normalsize\em 320 Osmond Lab}\\[\adressabstand]
        {\normalsize\em Penn State University}\\[\adressabstand]
        {\normalsize\em University Park, PA 16802}
        \\[-25\adressabstand]}      %
\date{November 10, 2002}
\maketitle   
\begin{abstract}
The properties of the space $\A$ of regular connections 
as a subset of the space $\Ab$ of generalized connections
in the Ashtekar framework are studied. For every choice of
compact structure group and smoothness category for the paths 
it is determined whether $\A$ is dense in $\Ab$ or not.
Moreover, it is proven that $\A$ has Ashtekar-Lewandowski measure zero
for every nontrivial structure group and every smoothness category.
The analogous results hold for gauge orbits instead of connections.
\end{abstract}

%------------------------------------------------------------------------%
%            Abschnitt: Introduction                                     %
%------------------------------------------------------------------------%
\section{Introduction}
One of the most important quantization methods is the functional 
integral approach. There, in a first step, one determines 
a physical Euclidian measure
on the configuration space and reconstructs then, in a second step,
the Hamiltonian theory using a kind of Osterwalder-Schrader procedure.
In the case of pure gauge field theories the configuration
space is the space $\ag$ of smooth connections (i.e.\ gauge fields) 
modulo smooth gauge transformations in some principal fibre bundle $P$ over
the (space[-time]) manifold $M$ with the structure group $\LG$.
However, in general, the structure of $\ag$ is very complicated: It is a
non-affine, non-compact, not finite-dimensional space and not a manifold.
Therefore when defining measures there, enormous problems appeared that
have been solved to date only partially. To avoid some of these problems,
Ashtekar, Isham and Lewandowski \cite{a72,a48,a30,a28} proposed to extend
the configuration space by distributional gauge orbits. Using 
$C^\ast$-algebraic techniques, this space $\agb$ could be interpreted 
as the compact spectrum of the $C^\ast$-algebra generated 
by the Wilson loops based on piecewise analytic paths.
Rendall \cite{e8} showed that $\ag$ can be densely imbedded into $\agb$.
This coincides fully with the expectations made in other rigorous 
functional integral approaches, e.g.\ in the Wiener-integral study
of the diffusion equation. 
Later on, also the spaces $\A$ and $\G$ of smooth connections and
gauge transformations, respectively, have been enlarged by
distributional objects, leading to the spaces $\Ab$ and $\Gb$. 
Using projective-limit techniques it has been shown that $\AbGb \iso \agb$.

However, there still had been a problem: In view of the desired 
applicability of the new approach to quantum gravity, due to its 
diffeomorphism invariance one should consider at least smooth paths
for the arguments of the parallel transports. The main problem
from the technical side is the following: In contrast to the 
piecewise analytical category two paths now can have infinitely many
intersection points without sharing just a complete interval. In other
words, two finite graphs need no longer be contained in a bigger third
graph being again finite. This problem has been first cured by
Baez and Sawin \cite{d3,d17} in the immersive smooth case 
using so-called webs. Recently \cite{paper2+4,paper3}, 
it has been shown that using so-called hyphs all smoothness categories
can be handled at the same footing, whereas webs and graphs now
are special kinds of hyphs. 

The knowledge about the r\^ole of regular (i.e.\ smooth) connections 
in these non-analytic frameworks, however, is still quite limited.
In the case of webs, only for the case of connected and semi-simple
structure groups $\LG$ it is known that $\A$ is dense in $\Ab$ (and
consequently $\ag$ in $\AbGb$ as well). Therefore, we will now study in this
article the properties of the spaces $\A$, $\G$ and $\ag$ viewed as 
subspaces of $\Ab$, $\Gb$ and $\AbGb$, respectively, in more detail
within the hyph framework.
The outline of this main body of the paper will be as follows: 
First we will discuss how these embeddings may depend on the 
necessary choice of a specific (typically non-smooth) trivialization of $P$.
It will turn out that all possible embeddings are in a 
certain sense equivalent. Afterwards, we will prove
that $\G$ is dense in $\Gb$ iff $\LG$ is connected. The corresponding
criterion for $\A$ and $\ag$, however, will be more difficult. As we will
see, whether the denseness is given or not, does crucially depend on
the structure group $\LG$ and on the used smoothness category for the 
paths. The most important result will be that in the non-analytic 
framework the denseness is at most be given for semi-simple $\LG$.
That section is followed by a discussion how to modify the
definitions of $\Ab$, $\Gb$ and $\AbGb$ to get possibly the desired denseness.
Finally, we will generalize the theorem of Marolf and Mour\~ao about
the Ashtekar-Lewandowski measure of $\A$ and $\ag$ to the 
case of general smoothness of paths.

We remark that an application of the denseness results of the present
article to the $C^\ast$-algebraic formulation of the Ashtekar framework
can be found in the paper \cite{e45} by Abbati and Mani\`a.

%------------------------------------------------------------------------%
%            Abschnitt: Preliminaries                                    %
%------------------------------------------------------------------------%
\section{Preliminaries}
In the section we briefly recall the basic definitions and conventions
used in this paper. General expositions can be found 
in \cite{a48,a30,a28} for the analytic framework and 
in \cite{diss,paper2+4,paper3} for arbitrary smoothness classes. The notion
and the properties of hyphs are discussed in \cite{paper3,diss}.

Let now $\LG$ be some arbitrary compact Lie group,
$M$ be a connected manifold having at least dimension $2$ and $m$ be
some arbitrary, but fixed point in $M$. (The restriction to $\dim M \geq 2$
is only due to technical reasons.) $P(M,\LG)$ -- or, shortly, $P$ -- denotes 
some principal fibre bundle $\pi: P \nach M$ with structure group $\LG$,
and $P_x$ denotes the fibre $\pi^{-1}{(x)}$ over $x\in M$ in $P$.
Next, we choose once and for all
some smoothness type $C^r$ for the paths
with $r\in\N_+$, $r=\infty$ (``smooth'') or 
$r = \omega$ (``analytic''), of course, with $r$ not being larger than the
smoothness category of $M$, and decide whether we will consider 
only immersive paths or also non-immersive paths. 
Now, $\Pf$ denotes the set of all (finite) paths in $M$. $\Pf$ is (after
imposing the standard equivalence relation, i.e., saying that
reparametrizations and insertions/deletions of retracings are irrelevant) 
a groupoid.
A graph is a finite set of edges (i.e.\ possibly closed, but 
elsewhere non-selfintersecting paths)
that intersect
each other at most in their endpoints. The subgroupoid generated
by the paths in a graph $\GR$ will be denoted by $\KG\GR$.
Graphs are ordered in the natural way: $\GR' \leq \GR''$ iff 
$\KG{\GR'} \teilmenge \KG{\GR''}$. The set $\Ab$ of generalized
connections $\qa$ is now defined by
\zglklein{\Ab := \varprojlim_\GR \Ab_\GR \iso \Hom(\Pf,\LG),}\noindent
with $\Ab_\gc := \Hom(\Pf_\gc,\LG) \iso \LG^{\elanz\gc}$ for all finite
sets $\gc$ of paths. (Often $\qa$ is written synonymously as 
$h_\qa$ to stress on the
homomorphy property.) Correspondingly, the set $\Gb$ of all 
generalized gauge transformations $\qg$ is defined by
\zglklein{\Gb := \varprojlim_\GR \Gb_\GR \iso \Maps(M,\LG),}\noindent
with $\Gb_\gc := \Maps(\Ver(\gc),\LG) \iso \LG^{\elanz\Ver(\gc)}$ 
for all finite $\gc\teilmenge\Pf$, where $\Ver(\gc)$ denotes
the set of all end points of the paths in $\gc$. The value of $\qg$ in
$x\in M$ is denoted by $\qg_x\in\LG$ or sometimes shortly by $g_x$.
The space $\Gb$ acts continuously on $\Ab$ via
\zglklein{h_{\qa\circ\qg}(\gamma) = 
              \qg_{\gamma(0)}^{-1} \: h_\qa(\gamma) \: \qg_{\gamma(1)}
              \:\:\:\: \text{ for all paths $\gamma$} }\noindent
yielding the factor space $\AbGb$ of generalized gauge orbits.
$\Ab$, $\Gb$ and $\AbGb$ are compact Hausdorff spaces.
Moreover, $\Gb$ acts on $\Gb$ continuously by conjugation:
$\qg \circ \qg' := \qg'{}^{-1} \cdot \qg \cdot \qg'$. 
This action is compatible with the action of $\Gb$ on $\Ab$, i.e.,
$(\qa \circ \qg) \circ \qg' = (\qa \circ \qg') \circ (\qg \circ \qg')$.
(Note, that the action in both cases is always from the right.)

A hyph $\hyph$ is a finite 
(ordered) set of edges $e_1,\ldots,e_{\elanz\hyph}$, 
where every $e_i \in \hyph$ possesses some ``free'' point. This means, 
for at least one direction none of the segments of $e_i$ 
starting in that point in this direction is a full segment
of some of the $e_1,\ldots,e_{i-1}$. The set of hyphs is ordered
analogously to the set of graphs. In contrast to the case of graphs,
this ordering in a direct ordering in the case of hyphs for {\em every}\/
smoothness category, i.e.,
for each two hyphs there is always some third hyph containing
both. Nevertheless, we have
\zglklein{\Ab \iso \varprojlim_\hyph \Ab_\hyph, \:\:\: \text{ } \:\:\:
          \Gb \iso \varprojlim_\hyph \Gb_\hyph \:\:\: \text{ and } \:\:\:
          \AbGb \iso \varprojlim_\hyph \Ab_\hyph/\Gb_\hyph.}\noindent
The corresponding continuous projections to the constituents of the projective
limits are in all these three cases denoted by $\pi_\hyph$, respectively.
It has been proven, that $\pi_\hyph$ is surjective for all hyphs $\hyph$. 

Finally, the Ashtekar-Lewandowski measure $\mu_0$ is the unique
regular Borel measure on $\Ab$ whose push-forward $(\pi_\hyph)_\ast\mu_0$ 
to $\Ab_\hyph$ coincides with the Haar measure there for every hyph $\hyph$. 
Since $\mu_0$ is 
$\Gb$-invariant, it can be seen as a measure on $\AbGb$ as well.

%------------------------------------------------------------------------%
%            Abschnitt: Embeddings                                       %
%------------------------------------------------------------------------%
\section{Embeddings}

In the projective-limit approach to Ashtekar connections one
needs a certain global trivialization of the underlying principal 
fibre bundle. That this trivialization can be chosen smooth 
is, of course, only possible if we are given a globally trivial bundle.
However, for Ashtekar connections we can take {\em any}\/ trivialization,
i.e., one for every fibre separately. It is, therefore, necessary to
investigate how the choice of trivializations influences the
embedding of the smooth objects into their extensions.
(Some investigations have been made already in \cite{a28}.)
But, as we will see this influence can
be neglected. More precisely, for any two trivializations we find
some isomorphism of $\Ab$ and $\Gb$, respectively, that maps the one embedding 
to the other and respects the action of $\Gb$ on $\Ab$. Consequently, we
will see that the embedding of $\ag$ into $\AbGb$ is even completely
independent of the choice of the embedding -- may the trivialization
be non-smooth everywhere. 

We start with
\bdf
Let 
$\triv = \{\triv_x : P_x \nach \LG 
               \mid x \in M\}$ 
be a set of fibre trivializations
and denote the parallel transports
according to $A\in\A$ along the path $\gamma\in\Pf$ by
$\partr_{\gamma,A} : P_{\gamma(0)} \nach P_{\gamma(1)}$.

The \df{embedding $\iota_\triv$ of the regular gauge theory} into the 
generalized gauge theory corresponding to $\triv$ consists of the
following three mappings all again denoted by $\iota_\triv$:
\bnum{3}
\item
$\iota_\triv : \A \nach \Ab$ with 
$h_{\iota_\triv(A)} (\gamma) 
 := \bigl(\triv_{\gamma(1)} \circ \partr_{\gamma,A} \circ \triv_{\gamma(0)}^{-1}\bigr)
        (e_\LG) \in \LG.$
\item
$\iota_\triv : \G \nach \Gb$ with 
$(\iota_\triv(\etr))(x) 
 := \bigl(\triv_{x} \circ \etr \circ \triv_{x}^{-1}\bigr)
        (e_\LG) \in \LG. $
\item
$\iota_\triv : \ag \nach \AbGb$ with 
$\iota_\triv([A]_\G) := [\iota_\triv(A)]_\Gb$.
\enum
\edf
Recall that $\triv_x$, in general, does not depend continuously on $x$.
\bprop
For every set $\triv$ of fibre trivializations, 
$\iota_\triv : \A\nach\Ab$ and $\iota_\triv : \G\nach\Gb$ 
and $\iota_\triv : \ag\nach\AbGb$  are
well-defined mappings. Moreover, the second one is a group homomorphism.
\eprop
\bpf
\bnum{3}
\item
$\iota_\triv : \A \nach \Ab$ is well-defined, since for all 
composable $\gamma, \delta \in \Pf$ we have
\bglklein
h_{\iota_\triv(A)} (\gamma \circ \delta) 
 & = & \bigl(
           \triv_{(\gamma\circ\delta)(1)} \circ 
           \partr_{\gamma\circ\delta,A} \circ 
           \triv_{(\gamma\circ\delta)(0)}^{-1}
       \bigr) (e_\LG) \\
 & = & \bigl(
           \triv_{\delta(1)} \circ 
           \partr_{\delta,A} \circ 
           \partr_{\gamma,A} \circ 
           \triv_{\gamma(0)}^{-1}
       \bigr) (e_\LG) \\
 & = & \bigl(
           \triv_{\delta(1)} \circ 
           \partr_{\delta,A} \circ 
           \triv_{\delta(0)}^{-1} \circ
           \triv_{\gamma(1)} \circ 
           \partr_{\gamma,A} \circ 
           \triv_{\gamma(0)}^{-1}
       \bigr) (e_\LG) \\
 & = & \bigl(
           \triv_{\gamma(1)} \circ 
           \partr_{\gamma,A} \circ 
           \triv_{\gamma(0)}^{-1}
       \bigr) (e_\LG) \cdot
       \bigl(
           \triv_{\delta(1)} \circ 
           \partr_{\delta,A} \circ 
           \triv_{\delta(0)}^{-1}
       \bigr) (e_\LG) \\ 
 & = & h_{\iota_\triv(A)} (\gamma) \: h_{\iota_\triv(A)} (\delta) 
\eglklein
by the invariance of the parallel transport under the action of $\LG$
on $P$. 
\item
$\iota_\triv : \G \nach \Gb$ is obviously well-defined,
and, moreover, a group homomorphism.
\item
$\iota_\triv : \ag \nach \AbGb$ is well-defined, since for all paths
$\gamma\in\Pf$ 
\bglklein
       h_{\iota_\triv(A \circ \etr)} (\gamma) 
 & = & \bigl(
            \triv_{\gamma(1)} \circ 
            \partr_{\gamma,A \circ \etr} \circ 
            \triv_{\gamma(0)}^{-1}
       \bigr)(e_\LG) \\
 & = & \bigl(
            \triv_{\gamma(1)} \circ 
            \etr \circ
            \partr_{\gamma,A} \circ 
            \etr^{-1} \circ
            \triv_{\gamma(0)}^{-1}
       \bigr)(e_\LG) \\
 & = & \bigl(
            \triv_{\gamma(1)} \circ 
            \etr \circ
            \triv_{\gamma(1)}^{-1} \circ
            \triv_{\gamma(1)} \circ 
            \partr_{\gamma,A} \circ 
            \triv_{\gamma(0)}^{-1} \circ 
{} \\ && \hspace{\fill} {} \circ
            \triv_{\gamma(0)} \circ 
            \etr^{-1} \circ
            \triv_{\gamma(0)}^{-1}
       \bigr)(e_\LG) \\
 & = & \bigl(
            \triv_{\gamma(0)} \circ 
            \etr^{-1} \circ
            \triv_{\gamma(0)}^{-1}
       \bigr) (e_\LG) \cdot
       \bigl(
            \triv_{\gamma(1)} \circ 
            \partr_{\gamma,A} \circ 
            \triv_{\gamma(0)}^{-1}
       \bigr) (e_\LG) \cdot 
{} \\ && \hspace{\fill} {} \cdot
       \bigl(
            \triv_{\gamma(1)} \circ 
            \etr \circ
            \triv_{\gamma(1)}^{-1}
       \bigr)(e_\LG) \\
 & = & (\iota_\triv(\etr))(\gamma(0))^{-1} \cdot 
       h_{\iota_\triv(A)} (\gamma) \cdot 
       (\iota_\triv(\etr))(\gamma(1)) \\
 & = & h_{\iota_\triv(A) \circ \iota_\triv(\etr)} (\gamma).
\eglklein
\qed
\enum
\epf

\bdf
\label{def:aequiv_triv}
Let $\triv_1$ and $\triv_2$ be two trivializations.

Then $\iota_{\triv_1}$ and $\iota_{\triv_2}$ are called \df{equivalent} iff
there is some $\qg_{\triv_2}^{\triv_1} \in \Gb$ such that
\bnum{2}
\item
$\iota_{\triv_2} (A) = \iota_{\triv_1} (A) \circ \qg_{\triv_2}^{\triv_1}$ 
for all $A\in\A$ and
\item
$\iota_{\triv_2} (\etr) = \iota_{\triv_1} (\etr) \circ \qg_{\triv_2}^{\triv_1}$ 
for all $\etr\in\G$.
\enum
\edf
\bcorr
If two trivializations are equivalent, then
the corresponding embeddings of $\ag$ coincide.
\ecorr
\bpf
By definition,
for two equivalent trivializations $\triv_1$ and $\triv_2$
there is some $\qg\in\Gb$ such that 
$\iota_{\triv_2}(A) = \iota_{\triv_1}(A) \circ \qg$,
hence
$\iota_{\triv_2}([A]_\G) \ident [\iota_{\triv_2}(A)]_\Gb
 = [\iota_{\triv_1}(A)]_\Gb \ident \iota_{\triv_1}([A]_\G)$
for all $A\in\A$.   
\qed
\epf
As it was to be expected, we have
\bthm
\label{thm:equiv(embed)}
All embeddings are mutually equivalent.
\ethm
\bpf
Let $\triv_1$ and $\triv_2$ be some trivializations.
Define $g_x := [((\triv_1)_x \circ (\triv_2)^{-1}_x)(e_\LG)]^{-1} \in \LG$
for $x\in M$ and set $\qg_{\triv_2}^{\triv_1} := (g_x)_{x\in M}$. 
Then we have for all $A\in\A$ and all $\gamma\in\Pf$
\bglklein
 &&
       h_{\iota_{\triv_1}(A) \circ \qg_{\triv_2}^{\triv_1}} (\gamma) 
\\
 & = & (\qg_{\triv_2}^{\triv_1})_{\gamma(0)}^{-1} \:
       h_{\iota_{\triv_1}(A)} (\gamma) \:
       (\qg_{\triv_2}^{\triv_1})_{\gamma(1)} \\
 & = & [((\triv_1)_{\gamma(0)} \circ (\triv_2)^{-1}_{\gamma(0)})(e_\LG)] \:
       \bigl(
            (\triv_1)_{\gamma(1)} \circ 
            \partr_{\gamma,A} \circ 
            (\triv_1)_{\gamma(0)}^{-1}
       \bigr)(e_\LG) \cdot
{} \\ && \hspace{\fill} {} \cdot
       [((\triv_1)_{\gamma(1)} \circ (\triv_2)^{-1}_{\gamma(1)})(e_\LG)]^{-1} \\
 & = & \bigl(
            (\triv_2)_{\gamma(1)} \circ 
            (\triv_1)_{\gamma(1)}^{-1} \circ
            (\triv_1)_{\gamma(1)} \circ 
            \partr_{\gamma,A} \circ 
            (\triv_1)_{\gamma(0)}^{-1} \circ
            (\triv_1)_{\gamma(0)} \circ 
            (\triv_2)_{\gamma(0)}^{-1}
       \bigr)(e_\LG)  \\
 & = & h_{\iota_{\triv_2}(A)} (\gamma),
\eglklein
hence $\iota_{\triv_2}(A) = \iota_{\triv_1}(A) \circ \qg_{\triv_2}^{\triv_1}$.
It is easy to see that analogously
$\iota_{\triv_2}(\etr) = \iota_{\triv_1}(\etr) \circ \qg_{\triv_2}^{\triv_1}$.
\qed
\epf

Until now we have not justified the notion ``embedding'' for
$\iota_\triv$. This will be caught up on next.
\bprop
For every set $\triv$ of fibre trivializations, 
$\iota_\triv : \A\nach\Ab$ and $\iota_\triv : \G\nach\Gb$ 
and $\iota_\triv : \ag\nach\AbGb$  
are injective. 
\eprop
\bpf
\bnum{3}
\item
Injectivity of $\iota_\triv : \A\nach\Ab$

For this, let $A_1$ and $A_2$ be two connections with
$h_{\iota_\triv(A_1)}(\gamma) = h_{\iota_\triv(A_2)}(\gamma)$ 
for all $\gamma\in\Pf$.
Assume $A_1 \neq A_2$. Then there is some $p\in P$ and some 
tangent vector $X\in T_p P$ such that $X$ is horizontal w.r.t.\ $A_1$,
but not w.r.t.\ $A_2$. If, however, $\gamma$ is some path through $\pi(p)$,
whose tangent vector in $\pi(p)$ equals $\pi_\ast X$, then the
horizontal lift $\widetilde\gamma_1$ of $\gamma$ w.r.t.\ $A_1$ 
has to be different from the horizontal lift $\widetilde\gamma_2$ w.r.t.\ $A_2$.
Hence, (at least along a suitable subpath of $\gamma$) 
the corresponding parallel transports have to be different as well, 
in contradition
to $h_{\iota_\triv(A_1)} = h_{\iota_\triv(A_2)}$ or
to the homeomorphy property of $\triv_x$ for some $x\in M$.

\item
Injectivity of $\iota_\triv : \G\nach\Gb$

This is obvious.

\item
Injectivity of $\iota_\triv : \ag\nach\AbGb$

Let $[A_1]$ and $[A_2]$ be in $\ag$ with 
$\iota_\triv([A_1]) = \iota_\triv([A_2])$. 
This means,
$\iota_\triv(A_2) = \iota_\triv(A_1) \circ \qg_\triv$ 
for some $\qg_\triv\in\Gb$ depending on $\triv$.
By Theorem \ref{thm:equiv(embed)} we can choose $\qg_\triv$ such that
$\qg_{\triv_2} = \qg_{\triv_1} \circ \qg_{\triv_2}^{\triv_1}$
with 
$(\qg_{\triv_2}^{\triv_1})_x := 
       [((\triv_1)_x \circ (\triv_2)^{-1}_x)(e_\LG)]^{-1}$
for every two trivializations $\triv_1$ and $\triv_2$.
We have to show that $\qg_\triv$ can be chosen regular, 
i.e.\ $\qg_\triv\in\iota_\triv(\G)$.

For this, let $\gamma(t)$ be some path in $M$ with initial point $x$.
The relation
$h_{\iota_\triv(A_2)}(\gamma) = 
       (\qg_\triv)_x^{-1} h_{\iota_\triv(A_1)}(\gamma) (\qg_\triv)_y$ 
for all $\gamma\in\pf{xy}$
guarantees that
$(\qg_\triv)_{\gamma(t)} 
 = h_{\iota_\triv(A_1)}(\gamma\einschr{[0,t]})^{-1} \:
   (\qg_\triv)_x \:
   h_{\iota_\triv(A_2)}(\gamma\einschr{[0,t]})$.
Since we reach every point in $M$ by some path $\gamma$ starting in $x$
and since $\triv$ is a bijection between $M \kreuz \LG$ and $\pi^{-1}(M) = P$,
there is a unique (not necessarily differentiable) 
gauge transform $\etr_\triv : P \nach P$ with
$(\iota_\triv(\etr_\triv))(y) 
 \ident \bigl(\triv_{y} \circ \etr_\triv \circ \triv_{y}^{-1}\bigr) (e_\LG)
 = (\qg_\triv)_y$ for all $y\in M$.

Now we assume that $\triv$ is $C^r$ over some open $U\teilmenge M$.
Then $h_{\iota_\triv(A)}(\gamma\einschr{[0,t]})$ 
depends differentiably on $t$ for every $\gamma\in\Pf$ with 
$\im\gamma\teilmenge U$ and, consequently,
$(g_\triv)_{\gamma(t)}$ is differentiable as well.
Using that $\triv$ is a diffeomorphism between
$U \kreuz \LG$ and $\pi^{-1}(U)\teilmenge P$ we see
that $\etr_\triv\einschr{\pi^{-1}(U)}$ is even differentiable.

Since around every point in $M$ there is some differentiable 
trivialization, we can define a global gauge transform
$\etr: P \nach P$ by $\etr(p) := \etr_\triv(p)$ for $p\in\pi^{-1}(U)$
if $\triv$ is any
differentiable trivialization over a neighbourhood $U$ of $x\in M$. 
Let now $\triv$ and $\triv'$ be two differentiable trivializations over $U$.
For $\gamma\in\Pf$ contained in $U$ and having base point $x$ we have
\bglklein
\iota_\triv(\etr_{\triv'}) (\gamma(t)) 
 & = & \bigl(\iota_{\triv'}(\etr_{\triv'}) \circ \qg^{\triv'}_{\triv}
       \bigr)(\gamma(t)) \\
 & = & (\qg^{\triv'}_{\triv})_{\gamma(t)}^{-1} \: 
          h_{\iota_{\triv'}(A_1)}(\gamma\einschr{[0,t]})^{-1} \:
          (\qg_{\triv'})_{x} \:
          h_{\iota_{\triv'}(A_2)}(\gamma\einschr{[0,t]}) \:
       (\qg^{\triv'}_{\triv})_{\gamma(t)} \\
 & = &  h_{\iota_{\triv}(A_1)}(\gamma\einschr{[0,t]})^{-1} \:
         (\qg^{\triv'}_{\triv})_{x}^{-1} \: 
         (\qg_{\triv'})_{x} \:
         (\qg^{\triv'}_{\triv})_{x} \:
        h_{\iota_{\triv}(A_2)}(\gamma\einschr{[0,t]}) \\
 & = &  h_{\iota_{\triv}(A_1)}(\gamma\einschr{[0,t]})^{-1} \:
         (\qg_{\triv})_{x} \:
        h_{\iota_{\triv}(A_2)}(\gamma\einschr{[0,t]}) \\
 & = & \iota_\triv(\etr_{\triv}) (\gamma(t)),
\eglklein
where we used again the transformation behaviour 
of $\iota_\triv$ and $\iota_{\triv'}$.
Since $\gamma$ was arbitrary in $U$ and $\iota_\triv$ is an embedding,
we get $\etr_\triv = \etr_{\triv'}$ on $\pi^{-1}(U)$. Hence, $\etr$ is
well-defined. Moreover, since every $\etr_\triv$ is differentiable in
the domain of differentiability of $\triv$, we get the differentiability
of $\etr$. The proof ends by $\iota_\triv(\etr) = \qg_\triv$.
\qed
\enum
\epf

Since we now know that the topological structure of the embeddings
of $\A$, $\G$ and $\ag$ as well as the structure of the action of
$\G$ on $\A$ is completely independent of the choice of the trivialization,
we will no longer worry about this and assume that we will have chosen 
in the next proofs, if necessary, silently some appropriate trivialization;
however, the results will be independent. In particular, we will simply
write $\A \teilmenge \Ab$ and $A$ instead of $\iota_\triv(A)$
etc.

%------------------------------------------------------------------------%
%            Abschnitt: Dichtheit                                        %
%------------------------------------------------------------------------%
\section{Denseness}
\label{uabschn:hyph:reg:dicht}
%------------------------------------------------------------------------%
%            Theorem: Eichtrfs. dicht <==> G zush.                       %
%------------------------------------------------------------------------%
In this main section we are going to study the properties
of the embeddings of the smooth objects into their Ashtekar extensions.
Before we come to the more difficult case of connections, we start
with the rather easy investigation, when the space of regular gauge
transformations is dense in that of generalized ones.
\bthm
$\G$ is dense in $\Gb$ iff $\LG$ is connected.
\ethm
\bpf 
\bhin
Let $\LG$ be not connected and let 
$g_1$ and $g_2$ be contained in the different
connected components
$\LG_{g_1}$ and $\LG_{g_2}$, respectively.
We define the open set 
$V := \LG_{g_1} \kreuz \LG_{g_2}$, choose 
a differentiable connected trivialization $U \teilmenge M$
and fix two points $x_1$ and $x_2$ in $U$. 
Then, for every edge $\gamma$ connecting $x_1$ and $x_2$ in $U$, 
the set $\pi^{-1}_\gamma(V) = (\pi_{x_1} \kreuz \pi_{x_2})^{-1}(V)$ 
is non-empty open in $\Gb$ and 
contains no regular gauge transform in $\G$.
\ehin  
\brueck
Let $\LG$ be connected. Then, obviously, for every hyph $\hyph$ and
every $\vec g\in\LG^{\elanz\Ver(\hyph)}$ there is
some regular gauge transform
$\eichtrf$ with $\eichtrf(x) = g_x$ for all $x\in\Ver(\hyph)$.
Hence $\pi_\hyph(\G) = \LG^{\elanz\Ver(\hyph)} = \Gb_\hyph$ and
Lemma \ref{lem:densecrit} yields the denseness.
\qed
\erueck
\epf   
%------------------------------------------------------------------------%
%            Theorem: Wann ist A dicht in Abar                           %
%------------------------------------------------------------------------%
The remaining part of this section is devoted to the proof of the following 
\bthm
$\A$ is dense in $\Ab$ precisely in the following cases: 
\bnum{3}
\item
in the analytic category for connected $\LG$;
\item
in the immersive $C^r$ category for connected and semisimple $\LG$;
\item
in the non-immersive $C^r$ category for trivial $\LG$.
\enum
The same criterion is true for the denseness of $\ag$ in $\AbGb$.
\ethm
%------------------------------------------------------------------------%
%            Lemma: G trivial                                            %
%------------------------------------------------------------------------%
We will prove this theorem case by case starting with the easiest one.
\blem
The denseness is given for trivial $\LG$.
\elem
\bpf
Obvious due to $\A = \Ab$ and $\ag = \AbGb$ for trivial $\LG$.
\qed
\epf
%------------------------------------------------------------------------%
%            Lemma: Analytisch oder Glatt und halbeinfach                %
%------------------------------------------------------------------------%
The next lemma contains the remaining cases of denseness. 
\blem
The denseness is given
\bnum{2}
\item
in the analytic category for connected $\LG$ and
\item
in the immersive $C^r$ category for connected and semisimple $\LG$;
\enum
\elem
\bpf
\bnum{2}
\item
In the analytic case the definition of $\Ab$, $\Gb$ and $\AbGb$ using
hyphs is equivalent to that using graphs \cite{paper3}.
Consequently, the denseness results for $\ag$ \cite{e8} 
and $\A$ \cite{diss} can be transferred immediately.
\item
In the smooth immersive case the definition of $\A$, $\G$ and $\ag$ is
equivalent to that using webs, where the denseness result
for $\A$ has been proven in \cite{e46}. Since even
$\pi_w(\A) = \LG^{\elanz w}$ for all webs $w$ \cite{e46}, we
get the denseness result by Corollary \ref{corr:densecrit} for $\ag$ 
as well.

Originally, webs have been defined only for smooth, i.e.\ $C^\infty$ paths.
However, one sees quite immediately, that this definition and the
corresponding subsequent theorems can be generalized to the case of arbitrary 
immersive $C^r$ paths ($r>0$). Therefore the denseness results can be
transferred as well. 
\qed
\enum
\epf
%------------------------------------------------------------------------%
%            Lemma: G nichtzush.                                         %
%------------------------------------------------------------------------%
Now let us turn to the cases of non-denseness again starting with the
two simplest ones.
\blem
The denseness is not given for non-connected $\LG$.
\elem
Note that using the standard identifications we have $\ag = \A/\Gb$.
\bpf
Let $\alpha$ be a closed edge contained completely 
in a contractible neighbourhood
of $m = \alpha(0) = \alpha(1)$ in $M$.
Then (using some trivialization being smooth there) we have  
$\pi_\alpha(\A) \teilmenge \LG_0$, where
$\LG_0\echteteilmenge\LG$ be the connected component of $e_\LG$.
On the other hand, since $\alpha$ is a hyph, 
by \cite{paper3} we have $\Ab_\alpha = \pi_\alpha(\Ab) = \LG$. 
The assertion now
follows from Lemma \ref{lem:densecrit}, because $\LG_0$ is, of course,
not dense in $\LG$.

\begin{figure}\begin{center}
\epsfig{figure=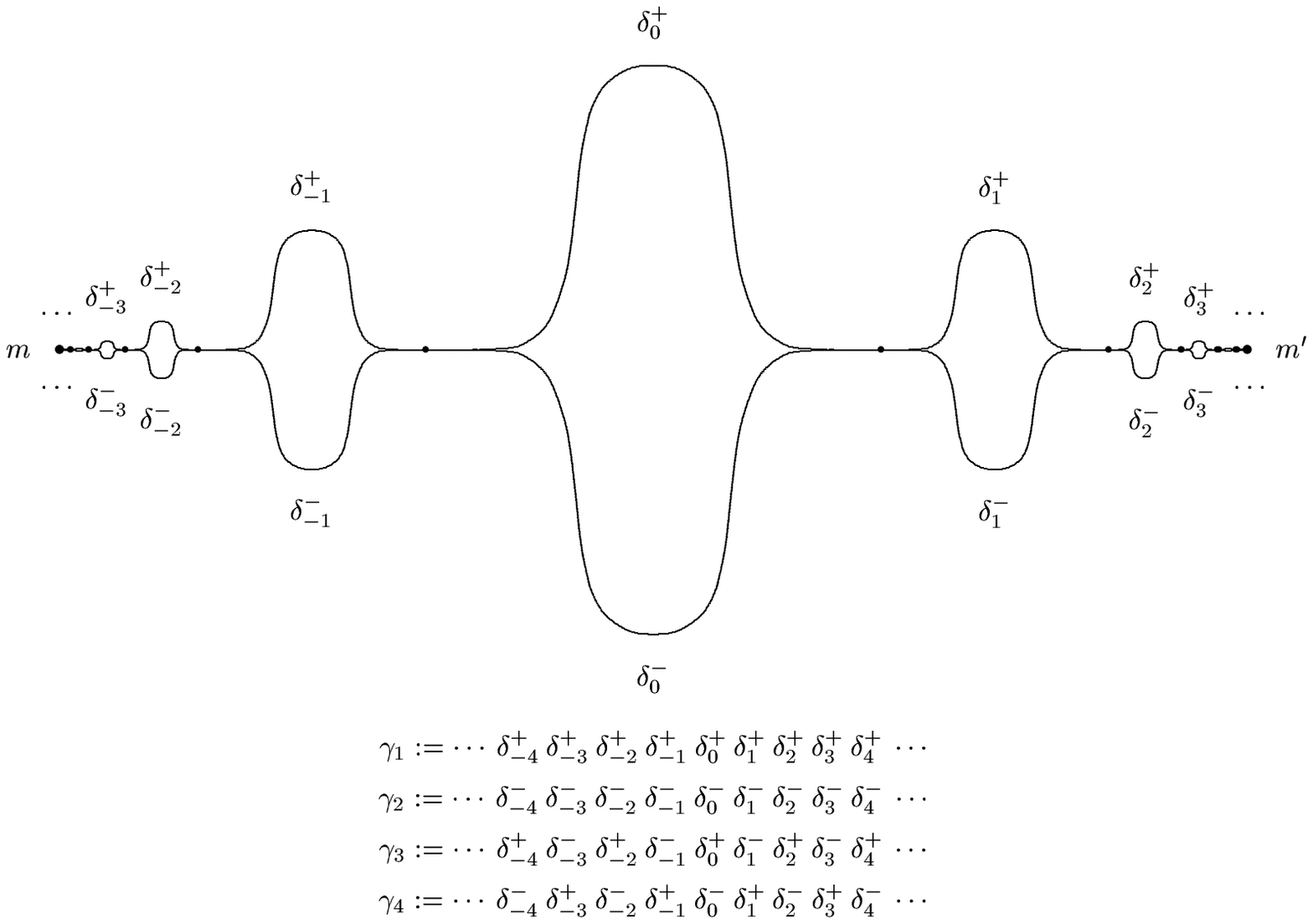,scale=0.85}
\caption{Paths used in the Proof of Lemma \ref{lem:nichtdicht_nichthalbeinf}}
\label{abb:wege_dichtheit}
\end{center}\end{figure}

The proof for $\ag$ now follows from Corollary \ref{corr:densecrit}:
The action of $\Gb_\alpha$ is just the adjoint action of $\LG$
which leaves $\LG_0$ invariant. 
\qed
\epf
%------------------------------------------------------------------------%
%            Lemma: nichtimmersiv                                        %
%------------------------------------------------------------------------%
\blem
The denseness is not given in the non-immersive smoothness category if
$\LG$ is non-trivial.
\elem
This lemma has been shown for $\A$ already in \cite{paper3}. We give
here a slightly modified proof that includes both $\A$ and $\ag$.
\bpf
Let $\gamma$ be a closed, immersive path without self-intersections
and $\gamma'(\tau):=\gamma(\tau^2)$. Then
$\gamma'$ is not equivalent to $\gamma$ (cf.\ \cite{paper2+4}) and
not an immersion. Moreover, $\hyph:=\{\gamma,\gamma'\}$ is a hyph.
However, since obviously $h_\gamma(A) = h_{\gamma'}(A)$ for all $A\in\A$,
we have $\pi_\hyph(\A) \teilmenge \{(g,g)|g\in\LG\}$ that is 
a non-dense subset of $\LG^2 = \Ab_\hyph$ for nontrivial $\LG$.
Lemma \ref{lem:densecrit} yields the assumption for $\A$,
Corollary \ref{corr:densecrit} that for $\ag$. In the last case
observe that $\Gb_\hyph$ is the adjoint action of $\LG$ which
leaves $\{(g,g)|g\in\LG\}$ invariant.
\qed
\epf
%------------------------------------------------------------------------%
%            Lemma: glatt und G nicht halbeinfach                        %
%------------------------------------------------------------------------%
Now we come to the most difficult case.
We will here reuse a certain example of paths 
(see Fig.\ \ref{abb:wege_dichtheit}) given 
in the paper \cite{d17} of Baez and Sawin.
It has been used there to show that the direct transfer of the
definition of spin networks from the analytic to the smooth category
is not possible. Here we will exploit another property of these 
paths: They are independent as graphs, but not holonomy-independent 
for abelian structure groups. This can be generalized to provide us with
\blem
\label{lem:nichtdicht_nichthalbeinf}
The denseness is not given for non-semisimple connected $\LG$ 
in the $C^r$ smoothness category.
\elem
\bpf
\bunum
\item
Let us consider the map
\fktdefabgesetzt{\spezabb}{\LG^4}{\LG.}{\vec g}{g_1 g_2 g_3^{-1} g_4^{-1}}\noindent
Since $\LG$ is supposed to be connected, it is isomorphic to 
$(\LG_\he \kreuz U(1)^k)/\LN$, where $\LG_\he$ is some
semisimple Lie group, $k$ is a natural number (by assumption $k\neq 0$)
and $\LN$ is a discrete central subgroup of $\LG_\he \kreuz U(1)^k$.
Now we define 
$K' := [\LG_\he \kreuz \{e\}]_\LN \teilmenge \LG$,
where $e$ is the identity in $U(1)^k$, and 
$K := \spezabb^{-1}(K') \teilmenge \LG^4$.
\bunum
\item
$K$ is $\Ad$-invariant.

Let $\vec g\in K$, i.e.\ $\spezabb(\vec g) = [g_\he,e]_\LN$ for some
$g_\he\in\LG_\he$.
Then for all $\widetilde g = [\widetilde g_\he, \widetilde g_\ab]_\LN\in\LG$
we have $\spezabb(\widetilde g^{-1} \vec g \widetilde g) = 
[\widetilde g_\he^{-1} g\ph_\he \widetilde g\ph_\he,e]_\LN\in K'$,
hence $\widetilde g^{-1} \vec g \widetilde g \in \spezabb^{-1}(K') = K$.
\item
$K$ is closed.

Obviously 
$\LG_\he \kreuz \{e\} \teilmenge \LG_\he \kreuz U(1)^k$ is closed, hence
$K' = [\LG_\he \kreuz \{e\}]_\LN \teilmenge \LG$ as well
by the compactness of $\LN$,
i.e.\ $K$ is closed by continuity of $\spezabb$.
\item
$K \neq \LG^4$.

Let us assume $K = \LG^4$.
Then we have $\spezabb(\vec g) \in K'$ for all $\vec g\in\LG^4$.
This means, we always get 
\bgl
&   & \spezabb([g_{1,\he},g_{1,\ab}]_\LN, \ldots, [g_{4,\he},g_{4,\ab}]_\LN) \\
& = & [g_{1,\he}\ph g_{2,\he}\ph g_{3,\he}^{-1} g_{4,\he}^{-1},
       g_{1,\ab}\ph g_{2,\ab}\ph g_{3,\ab}^{-1} g_{4,\ab}^{-1}]_\LN  \\
& = & [g_\he, e]_\LN,
\egl
\noindent
for some $g_\he\in\LG_\he$.
In particular, for all $\vec g_\ab\in (U(1)^k)^4$ there has to exist
some $n = (n_\he, n_\ab)\in\LN$, such that
$g_{1,\ab}\ph g_{2,\ab}\ph g_{3,\ab}^{-1} g_{4,\ab}^{-1} = n_\ab$.
Then, by the finiteness of $\LN$, also $U(1)^k$ has to be finite,
which requires $k = 0$. But, then $\LG$ is semisimple in contradiction
to our assumption.
\item
$K$ is not dense in $\LG^4$.

This follows simply from the fact that $K$ is a closed proper
subset of $\LG^4$.
\eunum
\item
Now we have a look at Fig.\ \ref{abb:wege_dichtheit}.
Let the paths $\gamma_i$ be given as indicated there, 
let $\gamma$ be some path from $m'$ to $m$, that does not intersect
any of the paths $\gamma_i$ and let $\alpha_i := \gamma_i  \: \gamma$.
Obviously, these four paths $\alpha_1$ till $\alpha_4$ form a hyph $\hyph$
with $m$ being the free point for every $\alpha_i$. 
However, although these paths are independent graph-theoretically, they are
{\em not}\/ independent w.r.t.\ regular connections:
Both $\gamma_1\gamma_2$ and $\gamma_4\gamma_3$ are paths,
that run precisely once through each $\delta_j^+$ and $\delta_j^-$ 
and precisely twice through $\gamma$.
Consequently, the abelian parts of 
$h_A(\gamma_1\gamma_2)$ and $h_A(\gamma_4\gamma_3)$ coincide for every
regular connection $A\in\A$ up to some $n_\ab$ 
in the abelian part of $\LN$.
Thus, 
\zgl{\spezabb(\pi_\hyph(A))
 = h_A(\gamma_1) h_A(\gamma_2) h_A(\gamma_3)^{-1} h_A(\gamma_4)^{-1}
 = [g_\he, n_\ab]_\LN = [g_\he n_\he^{-1}, e]_\LN \in K'}\noindent
for some $g_\he\in\LG_\he$ and some $n = (n_\he, n_\ab) \in \LN$,
hence 
$\pi_\hyph(\A) \teilmenge \spezabb^{-1}(K') = K$.

Since $K$ is not dense in $\LG^4 = \Ab_\hyph$, Lemma \ref{lem:densecrit}
implies that $\A$ is not dense in $\Ab$.
\item
The statement for $\ag$ follows now again by Corollary \ref{corr:densecrit},
since $\Gb_\hyph$ is the adjoint action of $\LG$ on $\LG^4$ leaving $K$ 
invariant.
\qed
\eunum
\epf

%------------------------------------------------------------------------%
%            Abschnitt: Discussion                                       %
%------------------------------------------------------------------------%
\section{Discussion}
The just proven lemma is in a certain sense a contrast to the usual
expectation, that within the functional integral framework
the classical theory should be a dense subset of the quantized theory
(cf.\ as an easiest example the Wiener integral for the diffusion
equation \cite{ReedSimon2}).
Let us therefore discuss how some modification of the definition of 
generalized connections could lead to the desired denseness of the regular
connections.
\bnum{3}
\item
The number of paths is reduced.

This can most easily be done by sharpening the assumptions regarding the 
smoothness of paths.
However, in view of applying the whole framework to quantum gravity,
at least $C^\infty$-paths should be allowed. Then, of course, we are no longer
able to couple theories to gravity which have non-semisimple structure groups.
\item
The number of independent paths is reduced.

One could try to plug the desired independence of paths into 
the equivalence relation of paths.
This idea closely follows the idea of holonomy equivalence used in the first 
articles on Ashtekar connections \cite{a72,a48}.
Here two paths are said to be equivalent iff they have
the same parallel transports for every regular connection.
It is obvious that this way the proof of 
Lemma \ref{lem:nichtdicht_nichthalbeinf} 
can no longer be used: 
For instance, $\alpha_1 \alpha_2$ and $\alpha_3 \alpha_4$
would be equivalent in the abelian case.
(But, note that this is not true in the nonabelian case.
Although here the holonomies are not arbitrarily selectable, they are not
always equal.)
However, the lack of denseness for nonabelian {\em and}\/ non-semisimple
structure groups remains true. This follows from the fact, that 
in the semisimple case 
$\pi_{\{\alpha_i\}_i}:\A\nach\LG^4$ is surjective, whence
$\alpha_1\alpha_2$ and $\alpha_3\alpha_4$ cannot be equivalent at least
if $\LG$ contains a nontrivial semisimple part.
In fact, it has been shown \cite{e46} that in the non-analytic immersed
category for non-abelian structure groups 
two paths are holonomically equivalent iff they are 
graph-theoretically equivalent. Therefore, in these cases the usage of
holonomy equivalence is not successful.
Moreover, this method has the drawback that in the case of non-immersed
paths it is probably very difficult to find at all concrete and explicit
criteria for independence of paths. 
\item
The range of the parallel transports is restricted.

This idea is precisely the basis for the definition of connections using
webs. Namely, here -- although done only for smooth and immersive paths --
the space $\Ab_\Web$ is not defined as a projective limit of all
$\Ab_w := \LG^{\elanz w}$, but as a projective limit of all images 
$\A_w := \pi_w(\A) \teilmenge \LG^{\elanz w}$ of {\em regular}\/
connections \cite{d3}. 
This way, automatically the surjectivity of $\pi_w:\A\nach\Ab_\Web$
is guaranteed, hence the denseness of $\A$ in $\Ab_\Web$ as well.
(The denseness follows, because first $\Ab_\Web = \varprojlim_w \A_w$, 
second the set of all webs is directed and third the projections
$\pi_w : \A \nach \A_w$ are per definitionem surjective. \cite{diss}  
The denseness of $\ag$ in $\Ab_\Web/\Gb$ is now trivial.)
\enum
Unfortunately, none of the three possibilities discussed above is free
of drawbacks such that a ``final'' decision about the definition of 
generalized connections can at most be given after studying more and
concrete physical models.

%------------------------------------------------------------------------%
%            Abschnitt: Measure                                          %
%------------------------------------------------------------------------%
\section{Measure}

In this final section we generalize the theorem of 
Marolf und Mour\~ao \cite{a42} about the 
Ashtekar-Lewandowski measure of $\A$ and $\ag$ 
to the case of arbitrary path categories considered here.
%------------------------------------------------------------------------%
%            Theorem: mu_0(A) = 0                                        %
%------------------------------------------------------------------------%
\bthm
\label{thm:mu0(ag)=0}
Both $\A$ and $\ag$ are contained in a set of Ashtekar-Lewandowski measure
$0$ provided $\LG$ is nontrivial.
\ethm
In the case that $\LG$ is trivial, we have $\A = \Ab$ and $\ag = \AbGb$
which means that the regular connection as well as the regular gauge orbits
form a set of full measure $1$.

Before we are going to prove the theorem, we note that 
Mour\~ao, Thiemann and Velhinho \cite{e21} were able to sharpen
the statement above in the case of $\Ab$ drastically:
Let $e$ be some edge and $e_s$ be the respective 
initial path of $e$ w.r.t.\ the interval $[0,s]$ for $s\in[0,1]$.
Moreover, let
\fktdefabgesetzt{q_\qa}{[0,1]}{\LG}{s}{h_\qa(e_s)}\noindent
for all $\qa\in\Ab$.
Then the set of all $\qa\in\Ab$ that possess just a single point in $[0,1]$
where $q_\qa$ is continuous, is contained in a $\mu_0$-zero subset.
This means, typically a generalized connection is nowhere continuous.
Although the proof has been done in the analytic case, it can be
transferred to the general case almost literally.
However, that proof does not give a statement on the measure of $\ag$
such that we will not reuse it.
Instead our proof is motivated by that of
Marolf and Mour\~ao:
The only accessible quantities within the Ashtekar framework are
parallel transports and holonomies.
Therefore one has to study how their behaviour is modified during the
transition from regular to generalized connections.
Typical for the regular case is -- in total contrast to the generalized
case -- that parallel transports depend in a certain sense continuously
on the paths. One can even prove that by means of a certain topology on
$\Pf$ the regular connections can be identified with the continuous 
homomorphisms from $\Pf$ to $M$ \cite{d20}. In particular,
``small'' paths have ``small'' parallel transports.
A more detailed analysis \cite{diss}
yields the well-known result (see, e.g., \cite{a42}) that for every
regular connection $A$ the difference 
$h_A(\alpha) - e_\LG$ is more or less proportional 
to the area enclosed by the sufficiently ``round'' loop $\alpha$.
This behaviour implies that the holonomies of a regular connection
are trapped for shrinking $\alpha$ in a small area around $e_\LG$
whose diameter decreases proportionally to
$\algnorm{h_A(\alpha) - e_\LG}$ and whose Haar measure consequently 
decreases as ${\FIL\alpha}^{\dim\LG}$.
However, generalized connections can even for very tiny $\alpha$
be anywhere in $\LG$.

Altogether we have the
%------------------------------------------------------------------------%
%            Beweis von Theorem \ref{thm:mu0(agb)=0}                     %
%------------------------------------------------------------------------%
\bpf
\bunum
\item
Let first $\dim\LG = 0$. Then (in an appropriate neighbourhood
of $m$) $h_\alpha(A) = e_\LG$ for all
regular $A$. Let now $(\alpha_i)_{i\in\N} \teilmenge \hg$
be a sequence of mutually non-intersecting closed edges having base point
$m$. Then
\zgl{\mu_0(\A) \leq \mu(\pi_{\{\alpha_1,\ldots,\alpha_i\}}^{-1}(\{e_\LG\}))
 = \mu_\Haar(\{e_\LG\})^i = (\elanz\LG)^{-i}}\noindent for all $i\in\N$.
Since $\LG$ is nontrivial, hence $\elanz\LG \geq 2 $, we have $\mu_0(\A) = 0$.
Analogously, we get $\mu(\ag) = 0$.
\item
Let now $\dim\LG > 0$.
We consider $\LG$ as a subset of some
$U(n) \teilmenge \Gl_\C(n) \teilmenge \C^{n\kreuz n}$ (and so 
$\Lieg\teilmenge \gl_\C(n) = \C^{n\kreuz n}$ as well),
choose some $\Ad \LG$-invariant
norm $\algnorm\cdot$ on $\C^{n\kreuz n}$ and define
$B_\varepsilon(e_\LG) := \{g\in\LG \mid \algnorm{g - e_\LG} < \varepsilon\}$ 
for all $\varepsilon\in\R_+$.
(For instance, 
$\algnorm{D} := \sup_{\vec x\in\C^n, \norm{\vec x} = 1} \norm{D\vec x}$,
$D \in \C^{n\kreuz n}$,
is $\Ad \LG$-invariant due to the unitarity of any compact group.)
Obviously, $B_\varepsilon(e_\LG)$ is always an $\Ad\LG$-invariant set.

Next, we choose some chart mapping
$\karte:M \obermenge U\nach\karte(U)$, such that $m \in U$.
W.l.o.g.\ $\karte(U) \teilmenge \R^{\dim M}$ be bounded and $\karte(m) = 0$.
We assign to $U$ the Euclidian metric and choose some surface $H\teilmenge U$
spanned by two coordinates.

Finally we assume that the chart image of every $\alpha\in\hg$ 
used below is a circle in $\karte(H) \teilmenge \R^2$. 
The area of the domain enclosed by $\alpha$ in $H$ be $\FIL\alpha$.
\item
Now we define for all $\alpha$ and all real $r\in\R_+$
the set
\zgl{U_{\alpha,r} := \pi_\alpha^{-1}(B_{r\FIL\alpha}(e_\LG)) \teilmenge \Ab.}\noindent 
being $\Gb$-invariant by the $\Ad$-invariance of $B_\varepsilon(e_\LG)$.
By the Appendix in \cite{paper6} we have
$\mu_0(U_{\alpha,r}) = \mu_\Haar(B_{r\FIL\alpha}(e_\LG))
                   \leq \const (r\FIL\alpha)^{\dim\LG}$.
Hence,
$\mu_0(U_{\alpha,r})$ goes to $0$, in particular, for 
$\FIL\alpha\konvrunter 0$.
\item
Let now $(\alpha_i)_{i\in\N}$ be some sequence of circles 
with $\FIL{\alpha_i}\konvrunter 0$,
such that each two of them have precisely $m$ as common point.
We define 
\zgl{U_r := \bigcap_{i\in\N} U_{\alpha_i,r}.}\noindent

Obviously
$\mu_0(U_r) \leq \inf_i \{\mu_0(U_{\alpha_i,r})\} = 0$.

\item
On the other hand, for every $A\in\A$ there is some $c_A\in\R_+$ with
$A\in U_{\alpha,c_{A}} \ident \pi_\alpha^{-1}(B_{c_{A}\FIL\alpha}(e_\LG))$
for all circles $\alpha$ (cf.\ Appendix in \cite{paper6}).
Hence, $A\in U_{c_{A}}$.
Therefore, $U := \bigcup_{r\in\N_+} U_r$ is obviously a 
$\mu_0$-zero subset containing $\A$. Since $U$ is even $\Gb$-invariant,
$U/\Gb$ is a $\mu_0$-zero subset as well, now containing 
$\A/\Gb = \ag$.
\qed
\eunum
\epf
We note finally, that the just proven support property is typical
for the description of physical theories in terms of functional integrals.
For instance, it is well-known that in the Wiener-integral case the
classical configuration space of all $C^1$-paths is a zero subset
in its completion (cf.\ \cite{ReedSimon2}). 
But, this is to be expected, since otherwise the measure on the
completion would define a non-trivial, physically ``reasonable'' 
measure on the classical configuration space as well, although the 
existence of such a measure usually is seen to be unlikely.

%------------------------------------------------------------------------%
%            Danksagung                                                  %
%------------------------------------------------------------------------%
\section*{Acknowledgements}
The author thanks Maria Cristina Abbati,
Abhay Ashtekar and Alessandro Mani{\`a} for
encouraging him to write this article.
The author has been supported in part by the Reimar-L\"ust-Stipendium
of the Max-Planck-Gesellschaft and by NSF grant PHY-0090091.

%------------------------------------------------------------------------%
%            Anhang                                                      %
%------------------------------------------------------------------------%
\anhangengl
\section{Denseness Criteria}
Let $A$ be a set and $\leq$ be a partial ordering on $A$. 
Next, let $X_a$ be a topological space for each $a\in A$ and
$\pi_{a_1}^{a_2}:X_{a_2}\nach X_{a_1}$ for all $a_1\leq a_2$ be a
continuous and surjective map with
$\pi_{a_1}^{a_2}\circ\pi_{a_2}^{a_3} = \pi_{a_1}^{a_3}$ if
$a_1\leq a_2\leq a_3$.
The corresponding projective limit $\varprojlim_{a\in A} X_{a}$
is denoted by $\pl X$.
Furthermore, let $\pi_a : \pl X \nach X_a$ be
the usual continuous 
projection on the $a$-component and $X$ be some subset of
$\pl X$. Finally, let $A$ be directed,
i.e., for any two $a',a''\in A$ there is an $a\in A$
with $a',a''\leq a$.
%------------------------------------------------------------------------%
%            Lemma: Dichtheitskrit.                                      %
%------------------------------------------------------------------------%
\blem   
\label{lem:densecrit}
$X$ is dense in $\pl X$ iff
$\pi_a(X)$ is dense in $X_a$ for all $a\in A$.
\elem
The $\invimpliz$-direction has already been proven in \cite{paper3}.
We quote it for completeness.
\bpf 
\bhin
Let $a\in A$ be arbitrary and let $U_a \teilmenge X_a$ be open.
Then $\pi_a^{-1}(U_a)$ is open in $\pl X$. Hence, there
is an $x\in X$ with $x\in\pi_a^{-1}(U_a)$. Consequently,
$\pi_a(x)\in\pi_a(\pi_a^{-1}(U_a)) \teilmenge U_a$.
\ehin
\brueck
Let $U\teilmenge\pl X$ be open and non-empty, i.e.\
$U \obermenge \bigcap_i \pi_{a_i}^{-1}(V_i)\neq\leeremenge$ with open
$V_i\teilmenge X_{a_i}$ and finitely many $a_i\in A$.
Since $A$ is directed, there is an $a\in A$ with $a_i\leq a$ for all $i$
and thus $U \obermenge \pi_a^{-1}\bigl(\bigcap_i
(\pi_{a_i}^a)^{-1}(V_i)\bigr)$
with non-empty $V:=\bigcap_i (\pi_{a_i}^a)^{-1}(V_i)\teilmenge X_a$.
$V$ is open because $\pi_{a_i}^a$ is continuous. 
Since $\pi_a(X)$ is dense in $X_a$,  
there is an $x\in X$ with $\pi_a(x)\in V$ and    
so $\pi_{a_i}(x)\in V_i$ for all $i$, hence $x\in U$.
\qed
\erueck
\epf
%------------------------------------------------------------------------%
%            Corollary: Dichtheitskrit.                                  %
%------------------------------------------------------------------------%
Now, let additionally $\pl G := \varprojlim_{a\in A} G_{a}$ be some
projective limit of compact topological groups $G_a$
acting continuously and compatibly on the corresponding
compact Hausdorff spaces $X_a$. Moreover, for both projective systems
all the projections $\pi_a$ be surjective.
For a precise definition of projective-limit group actions see
\cite{a30,a28,diss}.
\bcorr
\label{corr:densecrit}
$X/\pl G$ is dense in
$\pl X/\pl G$ iff $\pi_a(X)/G_a$ is dense in $X_a/G_a$ for all $a\in A$.

In particular, we have 
\bunum
\item
denseness if $\pi_a(X) = X_a$ for all $a\in A$
and 
\item
non-denseness if $\pi_a(X)$ is for all $a\in A$ contained 
in some $G_a$-invariant and non-dense subset of $X_a$. 
\eunum
\ecorr
\bpf
We define $\pl{X/G} := \varprojlim_{a\in A} X_a/G_a$. 
Then, by the assumptions, the map 
\fktdefabgesetzt{\abbAbGbagb}
                {\pl X/\pl G}{\pl{X/G}}{[(x_a)_{a\in A}]}{([x_a])_{a\in A}}\noindent
is a well-defined homeomorphism \cite{diss}. 
By the lemma above, $X/\pl G$ is now dense in $\pl X/\pl G$ iff
$\pi_a(\abbAbGbagb(X/\pl G)) = \pi_a(X)/G_a$ is dense in
$X_a/G_a$ for all $a\in A$.

To prove the non-denseness in the special case that $\pi_a(X)$ is for all
$a\in A$ contained in some $G_a$-invariant and non-dense subset of $X_a$,
use the fact that the canonical projections $X_a \nach X_a/G_a$ are always
open. The other special case is trivial.
\qed
\epf
The above assumptions for the projective limits are fulfilled
for $\pl X = \Ab$ and $\Gb$ defined using hyphs as label set $A$
\cite{paper2+4,paper3,diss}.

\leerezeile

\noindent
{\bf Keywords:} Smooth Connections, Ashtekar Connections, Gauge Orbits

\leerezeile

\noindent
{\bf MSC 2000:} 81T13 (Primary), 53C05 (Secondary)

\end{spacing}

\begin{thebibliography}{10}

\bibitem{e45}
{Maria Cristina Abbati and Alessandro Mani\`a: On the Spectrum of Holonomy
  Algebras. {\it J. Geom. Phys. (to appear)}. {\sf e-print: math-ph/0202004}.}

\bibitem{a72}
{Abhay Ashtekar and Chris J. Isham: Representations of the holonomy algebras of
  gravity and nonabelian gauge theories. {\it Class. Quant. Grav.} {\bf 9}
  (1992) {1433--1468}. {\sf e-print: hep-th/9202053}.}

\bibitem{a28}
{Abhay Ashtekar and Jerzy Lewandowski: Differential geometry on the space of
  connections via graphs and projective limits. {\it J. Geom. Phys.} {\bf 17}
  (1995) {191--230}. {\sf e-print: hep-th/9412073}.}

\bibitem{a30}
{Abhay Ashtekar and Jerzy Lewandowski: Projective techniques and functional
  integration for gauge theories. {\it J. Math. Phys.} {\bf 36} (1995)
  {2170--2191}. {\sf e-print: gr-qc/9411046}.}

\bibitem{a48}
{Abhay Ashtekar and Jerzy Lewandowski: Representation theory of analytic
  holonomy {$C^*$} algebras. In: {\it Knots and Quantum Gravity} (Riverside,
  CA, 1993), edited by John C. Baez, pp. 21--61, Oxford Lecture Series in
  Mathematics and its Applications~1 (Oxford University Press, Oxford, 1994).
  {\sf e-print: gr-qc/9311010}.}

\bibitem{d17}
{John C. Baez and Stephen Sawin: Diffeomorphism-invariant spin network states.
  {\it J. Funct. Anal.} {\bf 158} (1998) {253--266}. {\sf e-print:
  q-alg/9708005}.}

\bibitem{d3}
{John C. Baez and Stephen Sawin: Functional integration on spaces of
  connections. {\it J. Funct. Anal.} {\bf 150} (1997) {1--26}. {\sf e-print:
  q-alg/9507023}.}

\bibitem{paper3}
{Christian Fleischhack: Hyphs and the Ashtekar-Lewandowski Measure. {\it J.
  Geom. Phys. (to appear)}. \mbox{}MIS-Preprint 3/2000. {\sf e-print:
  math-ph/0001007}.}

\bibitem{diss}
{Christian Fleischhack: Mathematische und physikalische Aspekte
  verallgemeinerter Eichfeldtheorien im Ashtekarprogramm (Dissertation).
  Universit{\"a}t Leipzig, 2001.}

\bibitem{paper6}
{Christian Fleischhack: On the Support of Physical Measures in Gauge Theories.
  \mbox{}MIS-Preprint 70/2001. {\sf e-print: math-ph/0109030}.}

\bibitem{paper2+4}
{Christian Fleischhack: Stratification of the Generalized Gauge Orbit Space.
  {\it Commun. Math. Phys.} {\bf 214} (2000) {607--649}. {\sf e-print:
  math-ph/0001006, math-ph/0001008}.}

\bibitem{d20}
{Jerzy Lewandowski: Group of loops, holonomy maps, path bundle and path
  connection. {\it Class. Quant. Grav.} {\bf 10} (1993) {879--904}.}

\bibitem{e46}
{Jerzy Lewandowski and Thomas Thiemann: Diffeomorphism invariant quantum field
  theories of connections in terms of webs. {\it Class. Quant. Grav.} {\bf 16}
  (1999) {2299--2322}. {\sf e-print: gr-qc/9901015}.}

\bibitem{a42}
{Donald Marolf and Jos{\'e} M. Mour{\~a}o: On the support of the
  {A}shtekar-{L}ewandowski measure. {\it Commun. Math. Phys.} {\bf 170} (1995)
  {583--606}. {\sf e-print: hep-th/9403112}.}

\bibitem{e21}
{J. M. Mour{\~a}o, Th. Thiemann, and J. M. Velhinho: Physical properties of
  quantum field theory measures. {\it J. Math. Phys.} {\bf 40} (1999)
  {2337--2353}. {\sf e-print: hep-th/9711139}.}

\bibitem{ReedSimon2}
{Michael Reed and Barry Simon: {\it Methods of Modern Mathematical Physics,
  vol. 2 (Fourier Analysis, Self-Adjointness)}. Academic Press, Inc., London,
  1975.}

\bibitem{e8}
{Alan D. Rendall: Comment on a paper of {A}shtekar and {I}sham. {\it Class.
  Quant. Grav.} {\bf 10} (1993) {605--608}.}

\end{thebibliography}
\end{document}